\documentclass[%
 aip,
 jmp,%
 amsmath,amssymb,
 reprint,
groupedaddress,%
]{revtex4-1}

\usepackage{graphicx}
\usepackage{dcolumn}
\usepackage{bm}

\begin{document}

\preprint{aip /123-QED}

\title{Spin wave propagation and spin polarized electron transport in single crystal iron films }

\author{O. Gladii}
\altaffiliation{Univ. Grenoble Alpes, CEA, CNRS, SPINTEC, F-38000 Grenoble, France}
\author{D. Halley}
\author{Y. Henry}
\author{M. Bailleul}

\affiliation{
Institut de Physique et Chimie des Mat\'{e}riaux de Strasbourg, UMR 7504 CNRS, Universit\'{e} de Strasbourg, 23 rue du Loess, BP 43, 67034 Strasbourg Cedex 2, France}

\date{\today}

\begin{abstract}
 The technique of propagating spin wave spectroscopy is applied to a 20~nm thick Fe/MgO (001) film. The magnetic parameters extracted from the position of the resonance peaks are very close to those tabulated for bulk iron. From the propagating waveforms, a group velocity of 4~km/s and an attenuation length of about 6~$\mu$m are extracted for 1.6~$\mu$m-wavelength spin-wave at 18~GHz.  From the measured current-induced spin-wave Doppler shift, we also extract a surprisingly high degree of spin-polarization of the current of 83$\%$. This set of results makes single-crystalline iron a promising candidate for building devices utilizing high frequency spin-waves and spin-polarized currents.
\end{abstract}

\maketitle
\section{Introduction}

Future developments in the field of magnonics\cite{Kruglyak2010} require high-performance magnetic films, in which spin waves (SW) can propagate without suffering too much attenuation. The corresponding figure of merit is the attenuation length $L_{\text{att}}$, defined as the distance over which the spin-wave amplitude decays by a factor $e$, which depends on the magnetic parameters of the film (saturation magnetization, thickness, magneto-crystalline anisotropy) and on the magnetic precession losses.\cite{GurevichMelkov1996,StancilPrabhakar2009} More precisely, $L_{\text{att}}$ is inversely proportional to the effective magnetic damping parameter of the film $\alpha_{\text{eff}}$. The research in the field of magnonics is therefore concentrating on materials with low damping. One material of choice is Yttrium Iron Garnet (YIG), a ferrimagnetic insulator which exhibits damping factor in the range of $10^{-4}$, but which requires sophisticated growth technique, Liquid Phase Epitaxy for the relatively thick films used since the early days of spin-wave physics\cite{GurevichMelkov1996,StancilPrabhakar2009} or Pulsed Laser Deposition for the thinner films grown more recently.\cite{dallivykelly2013} Other drawbacks of YIG are its low saturation magnetization (about 0.18~T), which translates into relatively small group velocity and weak signals in inductive measurements, and the difficulty to interface it efficiently with standard spintronic elements such as spin-valves.

For these reasons, ferromagnetic metal films are being considered as a possible alternative. Indeed, their higher magnetic damping ($\alpha$ = 0.002 to 0.01) is compensated by their higher saturation magnetization (1 to 2~T) and their easier deposition and integration into spintronic stacks. Among the ferromagnetic metals considered so far, Permalloy ($\text{Ni}_{80}\text{Fe}_{20}$) has been used extensively: this very soft material, usually grown in polycrystalline form using standard magnetron sputtering, is magnetically very homogeneous and has moderate values of damping (0.008) and saturation magnetization (1~T). It was used successfully in the last years to build many different nanomagnonic devices.\cite{VlaminckBailleul2008,ChumakPirroSergaEtAl2009,demidov2011_taper} Two other classes of metal films were also explored by the magnonic community: (i) Heussler alloys\cite{Sebastian2012,zhu2011}, which are expected to have much smaller damping due to their half metallic character, but which remain quite delicate to obtain with very low values of damping, and (ii) CoFeB alloys,\cite{Yu2012,DemidovUrazhdinRinkevichEtAl2014} initially developed for tunnel magnetoresistance stacks, which exhibit a moderate damping of about 0.004 but whose magnetic properties depend quite strongly on the annealing conditions. We believe that many other ferromagnetic metals could be explored, so that a good compromise could be find for specific measurement/application conditions. A first step in this direction consists in exploring the potential of the three elemental transition metal ferromagnets. Nickel and cobalt do not seem to be adequate, because the former exhibits a relatively large damping of about 0.03 (Ref.~\onlinecite{bhagat1974}) and the latter has an hexagonal hcp structure quite difficult to handle. Iron on the other hand combines several advantages: a small damping in the bulk (about 0.002, see Refs.~\onlinecite{heinrich1966,bhagat1974,vanbockstal1990,cochran1991}, with similar values reported for single crystalline thin films\cite{urban2001,ScheckChengBailey2006}), a large saturation magnetization of 2.15~T, a bcc cubic structure allowing epitaxy over MgO\cite{yuasa2004} and GaAs\cite{brockmann1999}, and the possibility to include it in high performance TMR stacks.\cite{yuasa2004}

In this work, we explore in details spin-wave propagation in a pure iron film grown on MgO (001), taking advantage of the many possibilities offered by the technique of propagating spin-wave spectroscopy (PSWS). We start by describing the experiment (Sec.~\ref{sec_exp}). Then we discuss how the magnetic parameters of the film are deduced from the spin-wave resonance frequencies (Sec.\ref{sec_char_mag}) and how the SW propagation characteristics are extracted from the measured waveforms (Sec.~\ref{sec_char_prop}). Finally (Sec. \ref{sec_CISWDS}), we describe current-induced spin-wave Doppler shift measurements allowing one to determine the degree of spin-polarization of the current.

\section{Experiment}\label{sec_exp}

\begin{figure}
\centering
\includegraphics[width=7cm]{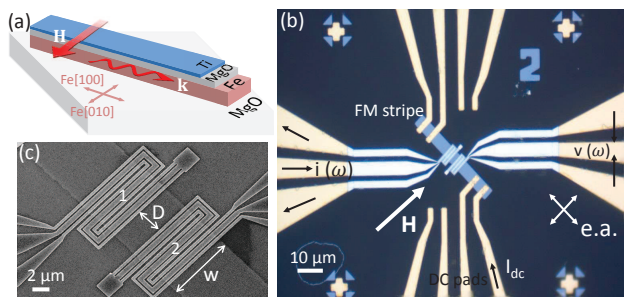}
\caption{(a) Sketch of the Fe/MgO (001) stack and of the measurement geometry. (b) Optical microscope image of a PSWS device. (c) Scanning electron microscope image of the spin-wave antennas.}.
\label{device}
\end{figure}

The Fe(001) film was grown by molecular beam epitaxy on a MgO(001) substrate. A 20~nm thick MgO buffer layer was first grown at 550$^{\circ}$C. Then, the t=20~nm Fe film was grown at 100$^{\circ}$C and annealed at 480$^{\circ}$C. Finally a MgO(8 nm)/Ti(4.5 nm) capping was deposited at room temperature. As usual for Fe/MgO(001) epitaxy, the [100] and [010] axis of Fe (which are easy axes for the cubic magneto-crystalline anisotropy) are rotated by 45$^{\circ}$ with respect to the [100] and [010] axis of MgO [Fig.\ref{device}(a)].
 
Fig.~\ref{device}(b) shows an optical microscope picture of an experimental device. It consists of a $w=10$~$\mu$m wide strip oriented along Fe[100], patterned from the continuous film by photolithography and ion milling, and a pair of spin-wave antennas fabricated by electron beam lithography and lift-off of Ti(10 nm)/Al(120 nm). In addition, the device comprises a $75$~nm $\text{SiO}_2$ layer isolating the antennas from the strip and a set of Ti(10~nm)/Au(60~nm) contacts, including a pair of coplanar waveguides for microwave connections to the antennas and four pads for DC connections to the strip. A scanning electron microscope picture of the spin-wave antennas is shown in Fig.~\ref{device}(c). The distribution of wave-vectors of the spin-waves generated by the antennas is governed by the Fourier transform of the spatial distribution of the microwave current, which contains a main peak at $k_\text M=3.8$~rad$/\mu$m and a secondary peak at $k_\text S=1.5$~rad$/\mu$m [see Fig.~8(b) in Ref.~\onlinecite{VlaminckBailleul2010}]. The external field $\textbf{H}$ is applied in the film plane, perpendicular to the strip, \textit{i.e.} parallel to the Fe[010] easy axis, so that the spin waves propagate in the so-called magnetostatic surface wave (MSSW) configuration.\cite{StancilPrabhakar2009}

The measurements are carried out by PSWS, the operating principle of which is explained in detail in Ref.~\onlinecite{VlaminckBailleul2010}. Briefly, when a microwave current $i(\omega)$ passes through one antenna it generates an oscillating magnetic field, therefore exciting spin waves which propagate in both directions along the strip. These spin wave induce an alternating voltage $v (\omega)$ both on the excitation antenna and on the second antenna separated from the first by a distance $D$. From the microwave reflection/transmission coefficients measured by a vector network analyzer connected to the two antennas $i,j=1,2$, we extract the inductance matrix $\Delta L_{ij}$, which is analyzed as follows. The self-inductances $\Delta L_{ii}$ exhibit absorption peaks associated with the excitation of spin waves. Following the position of these resonances as a function of the external field, one can derive the magnetic parameters of the film as with standard ferromagnetic resonance. The  mutual-inductances $\Delta L_{ij}$ exhibit oscillations. Following the amplitude and period of these oscillations as a function of $D$ allows one to extract the attenuation and the group velocity of the propagating spin waves.

\section{Determination of the magnetic parameters of the film}\label{sec_char_mag}

\begin{figure}
\includegraphics[width=6cm]{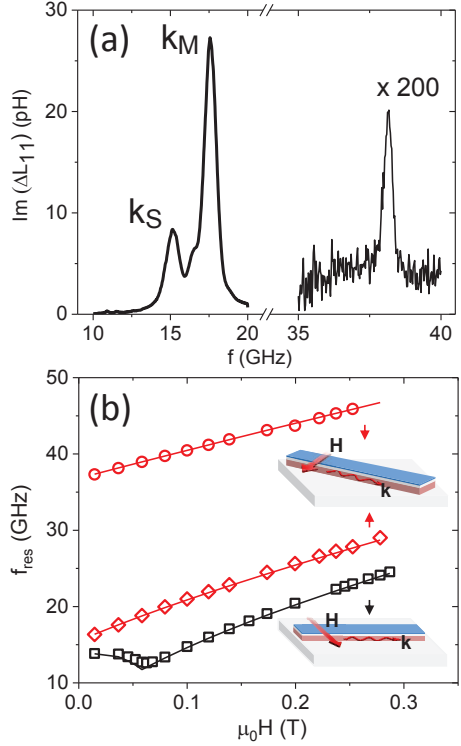}
\caption{(a) Self-inductance spectrum measured at $\mu_0H=36$~mT for the device of Fig.~\ref{device}. The peaks are identified as follows (in order of increasing frequency):  the secondary MSSW peak corresponding to a wave-vector $k_{\text S}=1.5$ rad$/\mu m$, the main MSSW peak corresponding to a wave-vector $k_{\text M}=3.8$ rad$/\mu m$ and the PSSW1 peak corresponding to a standing spin-wave mode across the film thickness. (b) Dependence on the external magnetic field of the frequency of the main MSSW peak (diamonds) and of the PSSW1 peak (circles) for the same device. The squares show the frequency of the main MSSW peak for a different device, where the Fe strip is oriented along a [110] hard axis of Fe (see the insets for the device geometries). The lines are the calculated mode frequencies for each mode.}
\label{reflexion}
\end{figure}

Fig.~\ref{reflexion}(a) shows the imaginary part of the self-inductance $\Delta L_{11}$, which corresponds to the power absorption, measured at $\mu_0H=36$~mT. One distinguishes two resonance peaks at 15~GHz and 18~GHz, which correspond to MSSWs with wave vectors $k_\text S$ and $k_\text M$, respectively. An additional peak of very low intensity is observed at about 38~GHz, which is attributed to the first perpendicular standing spin wave mode (PSSW1), a higher order mode for which the precession in the upper half of the film is in anti-phase with respect to that in its lower half. The maximum intensity of the main peak is about 27~pH, which corresponds to changes in impedance and reflection coefficient of 3~$\Omega$ and -30~dB, respectively. This is about 2.5 times larger than the intensity measured at the same frequency for a permalloy device with the same film thickness and lateral dimensions, thus confirming the advantage of using a material with a high saturation magnetization for such inductive measurements.

The variation of the frequency of the main MSSW peak as a function of the applied magnetic field is shown in Fig.~\ref{reflexion}(b) as diamonds. This variation is analyzed using the dispersion relation of MSSW:\cite{StancilPrabhakar2009}
\begin{multline}\label{dispersion}
f^{2}=(\frac{\mu_{0}\gamma}{2\pi})^{2}((H+H_{K})^{2}+(H+H_{K})M_{\text{eff}}+\\
+\frac{M_{\text{s}}M_{\text{eff}}}{4}(1-\exp(-2k_{M}t))),
\end{multline}
where $\gamma$ is the gyromagnetic ratio, $H_{K}=2K_1/(\mu_0M_{\text{s}})$ is the cubic anisotropy field, with $K_1$ the cubic anisotropy constant, $M_{\text{eff}}=M_{\text{s}}-H_{\text{u}}$ is the effective magnetization which takes into account an additional perpendicular uniaxial anisotropy $H_{\text{u}}$, and $t$ is the film thickness.\footnote{The standard expression of MSSW dispersion is modified to account for the cubic magnetic anisotropy of Fe and for a small perpendicular magnetic anisotropy of magneto-elastic and/or interface origin.} The solid line in Fig. \ref{reflexion}(b) was obtained by fitting the measured peak frequency to Eq.~\ref{dispersion}, which yields $\gamma/(2 \pi)=29$~GHz/T, $\mu_0 M_{\text{s}}=2.15$~T, $\mu_0 M_{\text{eff}}=2.08$~T and $\mu_0H_{\text K}=58$~mT, corresponding to $K_1=5\cdot10^{4}$~J/m$^3$. To determine the value of the exchange constant $A$, we use the PSSW1 peak, which in contrast to the MSSW peak is strongly influenced by the exchange interaction. Fitting the measured PSSW1 frequency [circles in Fig.~\ref{reflexion}(b)] to the Kalinikos-Slavin expression\cite{KalinikosN.1986} with the values of the other magnetic parameters given above yields a value $A=19$~pJ/m.

To illustrate directly the influence of the cubic anisotropy, we also use another device comprising a strip oriented along Fe[110] and measure it in the MSSW configuration (i.e. $\textbf{H}$ is applied along Fe[$1\overline{1}0$], which is an in-plane hard axis). The field dependence of the frequency of the main MSSW peak for this device [squares in Fig.~\ref{reflexion}(b)] is quite different from that measured on the [100] strip. Indeed, when increasing the field, the peak frequency field first decreases up to a field of about 58 mT  where it starts to increase, following the peak frequency of the [100] strip with a 116~mT shift towards higher field. The observed initial decrease of the peak frequency is attributed to the rotation of the magnetization vector $\textbf{M}$ toward the field direction. In such a hard axis geometry, the precession frequency is expected to reach its minimum at the value of field needed to align  $\textbf{M}$ along $\textbf{H}$, and this value is  equal  to the anisotropy field $H_\text K$. Determining numerically the angle of the equilibrium magnetization for each field, and calculating the corresponding MSSW frequency using the magnetic parameters indicated above [line in Fig.~\ref{dispersion}(b)], we obtain an excellent agreement with the experimental data, which confirms our determination of the cubic magnetic anisotropy. Note that all of the magnetic parameters determined in our film are in excellent agreement with those reported for bulk Fe.\cite{craikbook}

\section{Spin wave propagation characteristics}\label{sec_char_prop}

\begin{figure}
\includegraphics[width=8.5cm]{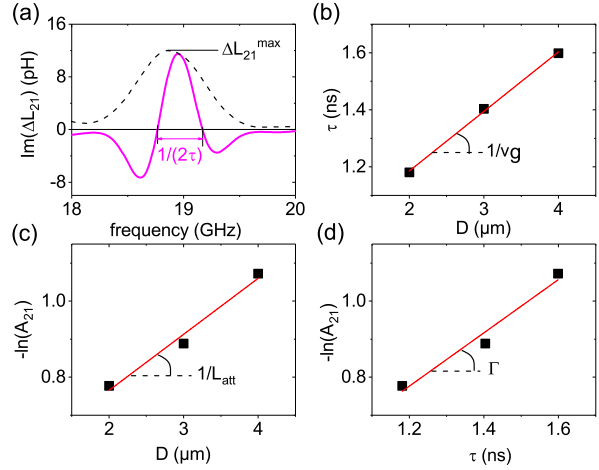}
\caption{(a) Measured mutual-inductance spectra for the device with an edge-to-edge distance $D=1 \mu m$ at a field $\mu_0H=58$~mT. (b) Dependence of the propagation time $\tau$ on the distance $D$. (c) Dependence of the logarithm of the normalized amplitude $A_{21}$ on the distance. (d) Dependence of the logarithm of $A_{21}$ on $\tau$.}
\label{transmission}
\end{figure}

Following the procedure in Ref.~\onlinecite{Gladii2016_PyPt}, we now analyze the mutual-inductance spectra in order to extract the characteristics of the propagating SW, namely their group velocity $v_{\text{g}}$, attenuation length $L_{\text{att}}$ and magnetization relaxation rate $\Gamma$. Figure~\ref{transmission}(a) shows the imaginary part (solid line) and the magnitude (dashed line) of the mutual inductance $\Delta L_{21}$ measured at $\mu_0H=58$~mT on a [100] strip with an edge-to-edge distance between antennas $D=1$~$\mu$m.  The measured waveform consists of an oscillation within an envelope having a peak shape. It can be characterized by two parameters: the period of the oscillation [the half of which is marked by a solid arrow in Fig.~\ref{transmission}(a)] and its maximum  amplitude $|\Delta L_{21}|^{\text {max}}$.

The period can be identified with the inverse of the group delay time $\tau=(D+D_0)/v_{\text{g}}$, where $D_0$ is an offset accounting for the finite-width of the antenna.\cite{ChangKostylevIvanovEtAl2014} Indeed, for two frequencies $f_1$ and $f_2$ close enough from each other, the difference of the propagation phase delays $\phi=-(D+D_0) k$ writes $\phi_2 - \phi_1 \backsimeq -2 \pi (f_2-f_1) \tau$ and amounts to $2 \pi$ for $f_2-f_1=1/\tau$. To evaluate the group velocity, we extract the group delay times for three devices with different $D$, perform a linear fit and identify the slope with $1/v_{\text{g}}$ [Fig.~\ref{transmission}(b)]. To evaluate the attenuation length, the amplitude of the mutual-inductance is first normalized to those of the self-inductances. This yields the quantity $A_{21}=|\Delta L_{21}|^{\text {max}}/\sqrt{|\Delta L_{11}|^{\text {max}}|\Delta L_{22}|^{\text {max}}}$, which is expected to decay exponentially with the distance as $A_{21}=exp (-(D+D_0)/L_{\text{att}})$. We therefore plot $-ln(A_{21})$ as a function of $D$, perform a linear fit and identify the slope with $1/L_{\text{att}}$ [Fig.~\ref{transmission}(c)]. Finally, one can also evaluate directly the spin-wave relaxation rate $\Gamma$, by plotting $-ln(A_{21})$ as a function of the measured $\tau$, the slope of the linear fit being identified with $\Gamma$ [Fig.~\ref{transmission}(d)]. This analysis yields the three SW propagation parameters related to each other through $L_{\text{att}}=v_{\text{g}}/\Gamma$. For a frequency of 18.9~GHz they amount to $v_{\text{g}}=4.8$~km/s, $L_{\text{att}}=6.8$~$\mu$m and $\Gamma=7 \cdot 10^8$~rad/s.

\begin{figure}[b]
\includegraphics[width=7cm]{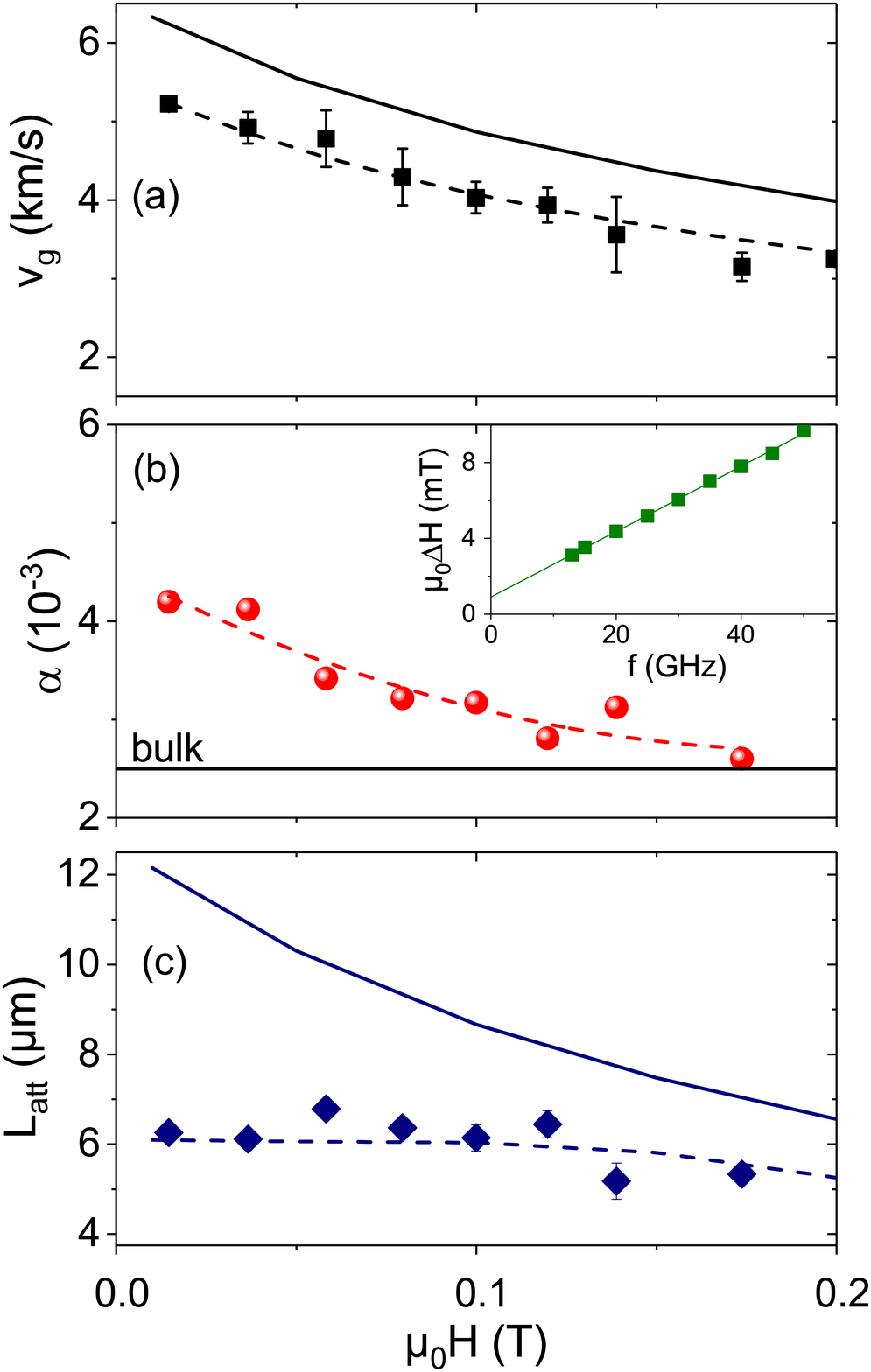}
\caption{Dependence of the group velocity (a), effective damping (b) and attenuation length (c) on the magnetic field. In each panel, the symbols are the measured data deduced using the method illustrated in Fig.~\ref{transmission} and the solid lines are the theoretical expectations. The dashed line in panels (a) and (b) are polynomial fit serving as guide for the eyes. The dashed line in panel (c) is deduced from these two fits using the relation $L_{\text{att}}=v_{\text g}/\Gamma$. The inset in panel (b) shows the frequency dependence of the FMR linewidth (full width at half maximum) measured on a similar film by broadband ferromagnetic resonance.}
\label{LattandDamp}
\end{figure}

In a next step, we follow the variation of the propagation characteristics as a function of the external field $H$ and compare them to theoretical predictions. Fig.~\ref{LattandDamp}(a) shows the measured and calculated values of the group velocity (squares and solid line, respectively). The latter is obtained by injecting the magnetic parameters given above in the expression derived from the MSSW relation dispersion [Eq.~(\ref{dispersion})], namely $v_{\text{g}}=\frac{\omega_{M}\omega_{\text{eff}}t}{4\omega}exp^{-2kt}$ with $\omega_{M}=\gamma \mu_{0}M_{\text{s}}$ and $\omega_{\text{eff}}=\gamma \mu_{0}M_{\text{eff}}$. The calculated $v_{\text{g}}$ are about 15$\%$ larger than the measured ones. We do not have an explanation for this difference yet, but we note that the overall agreement with the measurement is quite good. Fig.~\ref{LattandDamp}(b) shows the values of the effective damping $\alpha_{\text{eff}}$, estimated from the measured $\Gamma$ using the relation appropriate for an in-plane magnetized thin film,\cite{StancilPrabhakar2009} namely  $\Gamma=\alpha_{\text{eff}}(\omega_{0}+\omega_{M}/2)$ with $\omega_{0}=\gamma \mu_{0}(H+H_{K})$.  $\alpha_{\text{eff}}$ decreases significantly with increasing $H$  and reaches an asymptotic value of 0.0025 at high field. For comparison, the squares in the inset of Fig.~\ref{LattandDamp}(b) shows the frequency dependence of the resonance line-width $\Delta H$ measured on a similar Fe(20 nm)/MgO(001) film  by broadband ferromagnetic resonance. Fitting the data to the expression $\mu_0\Delta H=\mu_0\Delta H_0+\alpha\frac{4\pi f}{\gamma}$,\cite{GurevichMelkov1996} we extract an inhomogeneous broadening  $\mu_{0}\Delta H_0=0.9$~mT and $\alpha=0.0025$. The high field extrapolate of the SW effective damping is therefore in good agreement with the damping evaluated by FMR, which is also in line with the values measured in bulk Fe\cite{heinrich1966,bhagat1974,vanbockstal1990,cochran1991} and in epitaxial films\cite{urban2001,ScheckChengBailey2006}. We attribute the increase of effective damping observed at low field to the additional spin-wave relaxation related to inhomogeneous broadening.

Finally, the diamonds in Fig.~\ref{LattandDamp}(c) show the measured $L_{\text{att}}$, which has a nearly constant value of 6~$\mu$m. Not surprisingly, this is significantly smaller than the "ideal" attenuation length [dashed line in Fig.~\ref{LattandDamp}(c)] calculated using the theoretical expressions of $v_{\text{g}}$ and $\Gamma$ given above and a constant value of damping of 0.0025. From the analysis of panels (a) and (b), it is clear that this decrease arises both from a decrease of $v_g$ and a field dependent increase of $\alpha_{\text{eff}}$. Despite this deviation, the attenuation length keeps a pretty large value in comparison with other materials. Indeed, for the same SW wavelength and film thickness and for a frequency of 19~GHz, the attenuation length is about 1~$\mu$m in both permalloy ($\alpha=0.008$) and YIG ($\alpha=0.0004$). Note that, to reach such a SW frequency with these materials, the external field should be increased to 0.3 and 0.6~T, respectively.

\section{Current-induced spin-wave Doppler shift}\label{sec_CISWDS}

In order to investigate the spin-polarized electrical transport properties of the iron film, we now resort to the technique of current-induced spin-wave Doppler shift (CISWDS), injecting a DC current in the strip and following the changes of the PSWS waveform.\cite{VlaminckBailleul2008} The effect occurs due to  the interaction between the conduction electrons and the magnetization precession via the adiabatic spin transfer torque, which leads to a modification of the spin wave angular frequency taking the form of a Doppler shift  $\delta \omega_{\text{Dop}}=ku$. Here, $k$ is the SW wave-vector and $u$ is the effective spin drift velocity, which writes:
\begin{equation}\label{P}
u=P\frac{J}{e}\frac{\mu_{\text B}}{M_{\text s}},
\end{equation}
where $P=\frac{J_{\uparrow}-J_{\downarrow}}{J_{\uparrow}+J_{\downarrow}}$ is the degree of spin polarization of the electrical current, $J$ is the electrical current density, $e$ is the electron charge and $\mu_{\text B}$ is the Bohr magneton.

\begin{figure}
\includegraphics[width=8.5cm]{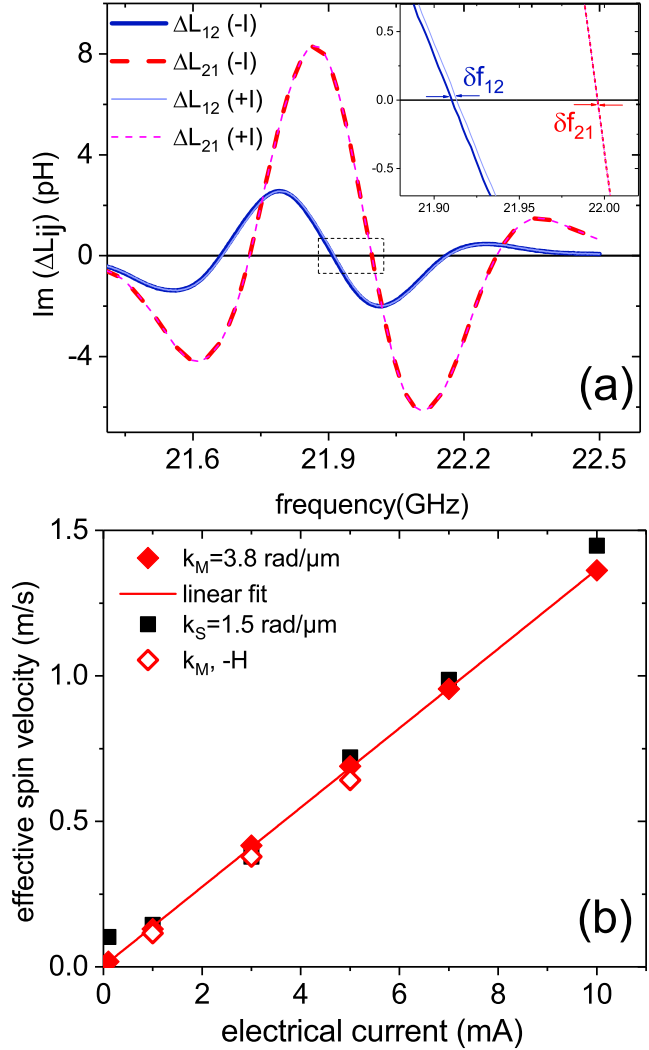}
\caption{(a) Mutual-inductance spectra measured on the D=1~$\mu$m device under an applied field $\mu_0H=$120~mT and a DC electrical current $I=\pm10$~mA. The imaginary parts of both $\Delta L_{21}$ ($k>0$) and $\Delta L_{12}$ ($k<0$) are shown for the two current polarities. The inset is a zoom of the waveforms allowing one to visualize the current-induced frequency shifts. (b) Current dependence of the effective spin velocity deduced from the part of the current-induced frequency shift which is odd in $k$. The red solid diamonds are for the main SW peak ($k_{\text{M}}=3.8$~rad/$\mu$m), the line is the corresponding linear fit. The black squares are for the secondary SW peak ($k_{\text{S}}=1.5$~rad/$\mu$m). The red open diamonds are for the main SW peak under an applied field $\mu_0H=-$120~mT.}
\label{doppler}
\end{figure}

Fig.~\ref{doppler}(a) shows the imaginary part of the mutual-inductance $\Delta L_{21}$ (dashed lines, which correspond to SW propagating from antenna 1 to antenna 2, \textit{i.e.} $k>0$) and $\Delta L_{12}$  (solid lines,  SW propagating from 2 to 1, \textit{i.e.} $k<0$). The data are measured on the device with D=1~$\mu$m and corresponds to the main MSSW peak.  A field $\mu_0H=$120~mT is applied along Fe[010] and the electrical current $I$ flows along Fe[100].  One first notices that the $\Delta L_{21}$ waveform is more intense and shifted towards higher frequency as compared to the $\Delta L_{12}$ one. This is due to the well-known non-reciprocal effects in the MSSW configuration. The amplitude non-reciprocity is attributed  to the circular polarization of the field produced by the antenna, which matches better the precession for one direction of propagation than for the other.\cite{SchneiderSergaNeumannEtAl2008} The frequency non-reciprocity is attributed to the combined effects of a non-reciprocal mode localization close to one film surface and an asymmetry of the  magnetic anisotropy at the two film surfaces.\cite{GladiiHaidarHenryEtAl2016}

Let us now concentrate on the current-induced SW frequency shift. The inset of Fig.~\ref{doppler}(a) shows a zoom allowing one to distinguish the frequency shift $\delta f_{12}$ between the propagating waveforms $\Delta L_{12}$ recorded at I = +10 mA (thin solid line) and I =-10 mA (thick solid line). Note that the frequency shift $\delta f_{21}$ between the two $\Delta L_{21}$ waveforms (thin and thick dashed lines) is not visible on this scale. Before analyzing these data one has to remove the influence of the Oersted field generated by the current by using the methods described in Refs.~\onlinecite{Haidar2013,HaidarBailleulKostylevEtAl2014}. Due to some asymmetries across the film thickness (\textit{e.g.} different probabilities of diffuse electron scattering on the two film surfaces), the Oersted field does not average strictly to zero. The resultant magnetic field adds to or subtracts from the external field and generates a current-induced SW frequency shift which is reciprocal (\textit{i.e.} it does not depend on the SW propagation direction). To eliminate this effect, we consider the quantity $\delta f_{\text{Dop}}=\delta f_{\text{odd}}=(\delta f_{21}-\delta f_{12})/4$.\cite{Haidar2013} Another effect is the so-called non-reciprocal Oersted field-induced frequency shift, which results from the combination of the antisymmetric distribution of the Oersted field with the MSSW modal profile non-reciprocal asymmetry which develops across thick enough films. Using the theory of Ref.~\onlinecite{HaidarBailleulKostylevEtAl2014} and the parameters of our Fe film, we calculate a shift of the order of 50~kHz for a current of 10~mA, which is much smaller than the measured frequency shift [$\delta f_{\text{Dop}}=770$~kHz from the data in Fig.\ref{doppler}(a)]. Solid diamonds in Fig.~\ref{doppler}(b) show the effective spin-velocity $u=2 \pi \delta f_{\text{Dop}}/k$, extracted as described above, and plotted as a function of the magnitude of the DC current $I$.  One recognizes a clear linear variation. From the slope of the linear fit, using Eq.~\ref{P} with the geometric and magnetic parameters of the Fe strip given above, we  obtain a degree of spin-polarization of the electrical current $P=0.83\pm0.05$. Note that very similar values of effective spin velocity are deduced from the current-induced frequency shift of the secondary MSSW peak [squares in Fig.~\ref{doppler}(b)], and also for an external field of opposite sign [open diamonds in Fig.~\ref{doppler}(b)], which is an indication that artifacts have been correctly accounted for by our extraction procedure.

The value $P=0.83$ we measure on our Fe film at room temperature (RT) can be compared to values determined using the same technique on different materials (P=0.5-0.7 for $\text{Ni}_{80}\text{Fe}_{20}$\cite{VlaminckBailleul2008,Zhu2010,Sekiguchi2012,Haidar2013,ChauleauBauerKoernerEtAl2014} and P=0.85-0.95 for $\text{CoFe}_{\text{1-x}}\text{Ge}_{\text{1-x}}$\cite{zhu2011}). In these two examples the large spin-polarizations are attributed to a strong spin-asymmetry of the electron scattering induced by the random alloy disorder, which could be understood from older works on impurity scattering in bulk ferromagnetic alloys\cite{CampbellFertchapter,Mertig1999}. In the case of a pure metal, such random alloy disorder scattering is absent and in our single-crystalline film, grain boundary scattering can also be ruled out. Diffuse electron scattering by the film surfaces is expected to have a small influence because the bulk electron mean free path, of the order of 2~nm,\cite{GurneySperiosuNozieresEtAl1993} is much smaller than the film thickness. Among the possible spin-dependent scattering processes listed in Ref.~\onlinecite{Haidar2013}, we are left only with the contributions of thermal disorder, namely those of phonons (thermal disorder of the lattice) and magnons (thermal disorder of the local magnetization). This is confirmed by temperature dependent resistivity measurements carried out on a similar Fe(20 nm)/MgO(001) film, which yield a RT value of 11~$\mu\Omega$.cm (similar to the one measured on the devices themselves, which is slightly larger than the 10~$\mu\Omega$.cm value tabulated for bulk Fe) and a 5~K resistivity of only 1.5~$\mu\Omega$.cm. Our result is a direct evidence of the strong spin-polarization of the electron scattering by thermal excitations in pure iron, which is a new finding. Indeed, deviations from the Matthiessen's rule measured in bulk dilute alloys indicated a sizeable spin-polarization of the resistivity for nickel\cite{FertCampbell1976} and cobalt\cite{LoegelGautier1971}, but such extraction could not be done unambiguously for iron. Fert and Campbell assumed the pure iron contribution to the resistivity to be non-polarized, based on the fact that the electron density of states for the two spin channels are quite similar at the Fermi level.\cite{FertCampbell1976}

On the contrary, we believe that in iron at room temperature, the phonon contribution to the electron resistivity is actually quite strongly spin-polarized, and that the magnon contribution, which has a tendency to depolarize the current because it is a spin-flip process, remains moderate. The latter statement is consistent with the findings of Raquet et al., who deduced a magnon contribution to the RT resistivity of the order of 2~$\mu\Omega$.cm from high field magnetoresistance measurements.\cite{RaquetViretSondergardEtAl2002} Because the inclusion of thermal excitations in ab-initio electron transport calculations remains problematic, there are not many theoretical works we can compare with. Liu \textit{et al.}\cite{LiuYuanWesselinkEtAl2015} and Ebert \textit{et al.}\cite{EbertMankovskyChadovaEtAl2015} used frozen disorder approaches to obtain RT phonon contributions which amount to of 5 and 9~$\mu\Omega$.cm, respectively. They could account for the 10~$\mu\Omega$.cm tabulated value by matching the thermal spin disorder to the measured temperature dependence of the saturation magnetization. Following another method, for a high enough temperature $T$, the phonon-contribution to the resistivity can be written $\rho=\frac{\pi \Omega_{\text{cell}} k_{\text B} T}{N \langle v_x^2 \rangle}\lambda_{\text{tr}}$, where $\Omega_{\text{cell}}$ is the unit cell volume, $k_{\text B}$ is the Boltzmann constant,  $N$ is the density of states at the Fermi Energy, $\langle v_x^2 \rangle$ is the average over the Fermi surface of the square of the component $v_x$ of the electron velocity along the electrical field, and $\lambda_{\text{tr}}$ is the transport electron-phonon coupling parameter, obtained by a suitable averaging of the electron-phonon spectral distribution functions determined by \textit{ab-initio} methods.\cite{SavrasovSavrasov1996} Using a realistic spin-polarized electron band-structure for Fe, Verstraete obtained $\lambda_{\text{tr}\uparrow}=0.1$ and $\lambda_{\text{tr}\downarrow}=0.2$,\cite{Verstraete2013} an asymmetry associated with the fact that, at the Fermi level of Fe, minority and majority electrons correspond to different parts of the d-band. This asymmetry, when combined with those of the density of states and Fermi velocity, namely $N_{x \uparrow}\langle v_{x}^2\rangle_{\uparrow}/N_{x \downarrow}\langle v_{\text x }^2\rangle_{\downarrow} \simeq 2$,\cite{ZhuZhangXuEtAl2008} results in a spin-polarization of the current of the order of 0.6, which is close to the value we measure. Interestingly, these two values are significantly higher than the value P=0.45 deduced from point-contact Andreev reflexion (PCAR) measurements on Fe films and foils.\cite{SoulenJr.1998} Indeed, the purely diffusive electrical transport regime probed in our room temperature CISWDS measurement is quite different from the partly ballistic regime probed in the low temperature PCAR technique. In the latter case, the spin-polarization is usually identified with that of $N\langle v_x \rangle$,\cite{SoulenJr.1998,ZhuZhangXuEtAl2008} which depends only on the electron band structure. In our case, the spin-dependence of the electron-phonon coupling plays an additional role.

\section{conclusion}
In this work we have shown that single-crystal iron constitutes a material of choice for propagating spin-waves. The large saturation magnetization translates into a relatively high group velocity and inductive signals of large amplitude. Although it is field-dependent and slightly higher than in bulk Fe, the effective damping remains small (0.0025-0.004). Iron is particularly promising for high-frequency operation because of the sizeable cubic magneto-crystalline anisotropy, which shifts the zero-field spin-wave frequency well above 10~GHz. In the last part of this paper, we have also shown that diffusive electron transport in iron at room temperature is strongly spin-polarized, which is attributed to a significant spin-asymmetry of the electron-phonon coupling. This indicates that, despite the absence of spin-polarized impurity or surface scattering, a pure material such as iron can be used efficiently for generating spin currents. The present study can be seen as a starting point for the investigation of several Fe-based alloys in the context for magnonic applications. Indeed, some of these alloys are  expected to exhibit much smaller damping\cite{ScheckChengBarsukovEtAl2007,devolder2013,SchoenThonigSchneiderEtAl2016} and larger spin-resistive signals\cite{ZahndVilaPhamEtAl2015} than the Ni-based alloys extensively used up to now.

\begin{acknowledgments}
The authors would like to thank F. Gautier, A. Fert, O. Bengone and H. Ebert for useful discussions, F. Abiza for the broadband FMR measurements, and the STnano facility and the Labex-NIE for giving us access to nanofabrication equipments. This work was supported by the Agence Nationale de la Recherche (France) under contract No. ANR-11-BS10-003 (NanoSWITI). O.G. thanks the Idex Unistra for doctoral funding.
\end{acknowledgments}


\begin{thebibliography}{48}%
\makeatletter
\providecommand \@ifxundefined [1]{%
 \@ifx{#1\undefined}
}%
\providecommand \@ifnum [1]{%
 \ifnum #1\expandafter \@firstoftwo
 \else \expandafter \@secondoftwo
 \fi
}%
\providecommand \@ifx [1]{%
 \ifx #1\expandafter \@firstoftwo
 \else \expandafter \@secondoftwo
 \fi
}%
\providecommand \natexlab [1]{#1}%
\providecommand \enquote  [1]{``#1''}%
\providecommand \bibnamefont  [1]{#1}%
\providecommand \bibfnamefont [1]{#1}%
\providecommand \citenamefont [1]{#1}%
\providecommand \href@noop [0]{\@secondoftwo}%
\providecommand \href [0]{\begingroup \@sanitize@url \@href}%
\providecommand \@href[1]{\@@startlink{#1}\@@href}%
\providecommand \@@href[1]{\endgroup#1\@@endlink}%
\providecommand \@sanitize@url [0]{\catcode `\\12\catcode `\$12\catcode
  `\&12\catcode `\#12\catcode `\^12\catcode `\_12\catcode `\%12\relax}%
\providecommand \@@startlink[1]{}%
\providecommand \@@endlink[0]{}%
\providecommand \url  [0]{\begingroup\@sanitize@url \@url }%
\providecommand \@url [1]{\endgroup\@href {#1}{\urlprefix }}%
\providecommand \urlprefix  [0]{URL }%
\providecommand \Eprint [0]{\href }%
\providecommand \doibase [0]{http://dx.doi.org/}%
\providecommand \selectlanguage [0]{\@gobble}%
\providecommand \bibinfo  [0]{\@secondoftwo}%
\providecommand \bibfield  [0]{\@secondoftwo}%
\providecommand \translation [1]{[#1]}%
\providecommand \BibitemOpen [0]{}%
\providecommand \bibitemStop [0]{}%
\providecommand \bibitemNoStop [0]{.\EOS\space}%
\providecommand \EOS [0]{\spacefactor3000\relax}%
\providecommand \BibitemShut  [1]{\csname bibitem#1\endcsname}%
\let\auto@bib@innerbib\@empty
\bibitem [{\citenamefont {Kruglyak}, \citenamefont {Demokritov},\ and\
  \citenamefont {Grundler}(2010)}]{Kruglyak2010}%
  \BibitemOpen
  \bibfield  {author} {\bibinfo {author} {\bibfnamefont {V.~V.}\ \bibnamefont
  {Kruglyak}}, \bibinfo {author} {\bibfnamefont {S.~O.}\ \bibnamefont
  {Demokritov}}, \ and\ \bibinfo {author} {\bibfnamefont {D.}~\bibnamefont
  {Grundler}},\ }\bibfield  {title} {\enquote {\bibinfo {title} {Magnonics},}\
  }\href {\doibase 10.1088/0022-3727/43/26/264001} {\bibfield  {journal}
  {\bibinfo  {journal} {Journal of Physics D: Applied Physics}\ }\textbf
  {\bibinfo {volume} {43}},\ \bibinfo {pages} {264001} (\bibinfo {year}
  {2010})}\BibitemShut {NoStop}%
\bibitem [{\citenamefont {Gurevich}\ and\ \citenamefont
  {Melkov}(1996)}]{GurevichMelkov1996}%
  \BibitemOpen
  \bibfield  {author} {\bibinfo {author} {\bibfnamefont {A.~G.}\ \bibnamefont
  {Gurevich}}\ and\ \bibinfo {author} {\bibfnamefont {G.}~\bibnamefont
  {Melkov}},\ }\href@noop {} {\emph {\bibinfo {title} {Magnetization
  oscillations and waves}}},\ edited by\ \bibinfo {editor} {\bibfnamefont
  {A.~G.}\ \bibnamefont {Gurevich}}\ and\ \bibinfo {editor} {\bibfnamefont
  {G.}~\bibnamefont {Melkov}}\ (\bibinfo  {publisher} {CRC Press, Boca Raton},\
  \bibinfo {year} {1996})\BibitemShut {NoStop}%
\bibitem [{\citenamefont {Stancil}\ and\ \citenamefont
  {Prabhakar}(2009)}]{StancilPrabhakar2009}%
  \BibitemOpen
  \bibfield  {author} {\bibinfo {author} {\bibfnamefont {D.~D.}\ \bibnamefont
  {Stancil}}\ and\ \bibinfo {author} {\bibfnamefont {A.}~\bibnamefont
  {Prabhakar}},\ }\href {\doibase 10.1007/978-0-387-77865-5} {\emph {\bibinfo
  {title} {Spin Waves. Theory and applications}}},\ \bibinfo {edition} {1st}\
  ed.,\ edited by\ \bibinfo {editor} {\bibfnamefont {D.~D.}\ \bibnamefont
  {Stancil}}\ and\ \bibinfo {editor} {\bibfnamefont {A.}~\bibnamefont
  {Prabhakar}}\ (\bibinfo  {publisher} {Springer US},\ \bibinfo {year}
  {2009})\BibitemShut {NoStop}%
\bibitem [{\citenamefont {d'Allivy Kelly}\ \emph {et~al.}(2013)\citenamefont
  {d'Allivy Kelly}, \citenamefont {Anane}, \citenamefont {Bernard},
  \citenamefont {Youssef}, \citenamefont {Hahn}, \citenamefont {Molpeceres},
  \citenamefont {Carretero}, \citenamefont {Jacquet}, \citenamefont {Deranlot},
  \citenamefont {Bortolotti}, \citenamefont {Lebourgeois}, \citenamefont
  {Mage}, \citenamefont {de~Loubens}, \citenamefont {Klein}, \citenamefont
  {Cros},\ and\ \citenamefont {Fert}}]{dallivykelly2013}%
  \BibitemOpen
  \bibfield  {author} {\bibinfo {author} {\bibfnamefont {O.}~\bibnamefont
  {d'Allivy Kelly}}, \bibinfo {author} {\bibfnamefont {A.}~\bibnamefont
  {Anane}}, \bibinfo {author} {\bibfnamefont {R.}~\bibnamefont {Bernard}},
  \bibinfo {author} {\bibfnamefont {J.~B.}\ \bibnamefont {Youssef}}, \bibinfo
  {author} {\bibfnamefont {C.}~\bibnamefont {Hahn}}, \bibinfo {author}
  {\bibfnamefont {A.~H.}\ \bibnamefont {Molpeceres}}, \bibinfo {author}
  {\bibfnamefont {C.}~\bibnamefont {Carretero}}, \bibinfo {author}
  {\bibfnamefont {E.}~\bibnamefont {Jacquet}}, \bibinfo {author} {\bibfnamefont
  {C.}~\bibnamefont {Deranlot}}, \bibinfo {author} {\bibfnamefont
  {P.}~\bibnamefont {Bortolotti}}, \bibinfo {author} {\bibfnamefont
  {R.}~\bibnamefont {Lebourgeois}}, \bibinfo {author} {\bibfnamefont {J.-C.}\
  \bibnamefont {Mage}}, \bibinfo {author} {\bibfnamefont {G.}~\bibnamefont
  {de~Loubens}}, \bibinfo {author} {\bibfnamefont {O.}~\bibnamefont {Klein}},
  \bibinfo {author} {\bibfnamefont {V.}~\bibnamefont {Cros}}, \ and\ \bibinfo
  {author} {\bibfnamefont {A.}~\bibnamefont {Fert}},\ }\bibfield  {title}
  {\enquote {\bibinfo {title} {Inverse spin hall effect in nanometer-thick
  yttrium iron garnet/pt system},}\ }\href {\doibase 10.1063/1.4819157}
  {\bibfield  {journal} {\bibinfo  {journal} {Appl. Phys. Lett.}\ }\textbf
  {\bibinfo {volume} {103}},\ \bibinfo {pages} {082408} (\bibinfo {year}
  {2013})}\BibitemShut {NoStop}%
\bibitem [{\citenamefont {Vlaminck}\ and\ \citenamefont
  {Bailleul}(2008)}]{VlaminckBailleul2008}%
  \BibitemOpen
  \bibfield  {author} {\bibinfo {author} {\bibfnamefont {V.}~\bibnamefont
  {Vlaminck}}\ and\ \bibinfo {author} {\bibfnamefont {M.}~\bibnamefont
  {Bailleul}},\ }\bibfield  {title} {\enquote {\bibinfo {title}
  {Current-induced spin-wave doppler shift},}\ }\href {\doibase
  10.1126/science.1162843} {\bibfield  {journal} {\bibinfo  {journal}
  {Science}\ }\textbf {\bibinfo {volume} {322}},\ \bibinfo {pages} {410--413}
  (\bibinfo {year} {2008})}\BibitemShut {NoStop}%
\bibitem [{\citenamefont {Chumak}\ \emph {et~al.}(2009)\citenamefont {Chumak},
  \citenamefont {Pirro}, \citenamefont {Serga}, \citenamefont {Kostylev},
  \citenamefont {Stamps}, \citenamefont {Schultheiss}, \citenamefont {Vogt},
  \citenamefont {Hermsdoerfer}, \citenamefont {Laegel}, \citenamefont {Beck},\
  and\ \citenamefont {Hillebrands}}]{ChumakPirroSergaEtAl2009}%
  \BibitemOpen
  \bibfield  {author} {\bibinfo {author} {\bibfnamefont {A.~V.}\ \bibnamefont
  {Chumak}}, \bibinfo {author} {\bibfnamefont {P.}~\bibnamefont {Pirro}},
  \bibinfo {author} {\bibfnamefont {A.~A.}\ \bibnamefont {Serga}}, \bibinfo
  {author} {\bibfnamefont {M.~P.}\ \bibnamefont {Kostylev}}, \bibinfo {author}
  {\bibfnamefont {R.~L.}\ \bibnamefont {Stamps}}, \bibinfo {author}
  {\bibfnamefont {H.}~\bibnamefont {Schultheiss}}, \bibinfo {author}
  {\bibfnamefont {K.}~\bibnamefont {Vogt}}, \bibinfo {author} {\bibfnamefont
  {S.~J.}\ \bibnamefont {Hermsdoerfer}}, \bibinfo {author} {\bibfnamefont
  {B.}~\bibnamefont {Laegel}}, \bibinfo {author} {\bibfnamefont {P.~A.}\
  \bibnamefont {Beck}}, \ and\ \bibinfo {author} {\bibfnamefont
  {B.}~\bibnamefont {Hillebrands}},\ }\bibfield  {title} {\enquote {\bibinfo
  {title} {Spin-wave propagation in a microstructured magnonic crystal},}\
  }\href {\doibase 10.1063/1.3279138} {\bibfield  {journal} {\bibinfo
  {journal} {Appl. Phys. Lett.}\ }\textbf {\bibinfo {volume} {95}},\ \bibinfo
  {pages} {262508} (\bibinfo {year} {2009})}\BibitemShut {NoStop}%
\bibitem [{\citenamefont {Demidov}\ \emph {et~al.}(2011)\citenamefont
  {Demidov}, \citenamefont {Kostylev}, \citenamefont {Rott}, \citenamefont
  {Münchenberger}, \citenamefont {Reiss},\ and\ \citenamefont
  {Demokritov}}]{demidov2011_taper}%
  \BibitemOpen
  \bibfield  {author} {\bibinfo {author} {\bibfnamefont {V.~E.}\ \bibnamefont
  {Demidov}}, \bibinfo {author} {\bibfnamefont {M.~P.}\ \bibnamefont
  {Kostylev}}, \bibinfo {author} {\bibfnamefont {K.}~\bibnamefont {Rott}},
  \bibinfo {author} {\bibfnamefont {J.}~\bibnamefont {Münchenberger}},
  \bibinfo {author} {\bibfnamefont {G.}~\bibnamefont {Reiss}}, \ and\ \bibinfo
  {author} {\bibfnamefont {S.~O.}\ \bibnamefont {Demokritov}},\ }\bibfield
  {title} {\enquote {\bibinfo {title} {Excitation of short-wavelength spin
  waves in magnonic waveguides},}\ }\href {\doibase 10.1063/1.3631756}
  {\bibfield  {journal} {\bibinfo  {journal} {Appl. Phys. Lett.}\ }\textbf
  {\bibinfo {volume} {99}},\ \bibinfo {pages} {082507} (\bibinfo {year}
  {2011})}\BibitemShut {NoStop}%
\bibitem [{\citenamefont {Sebastian}\ \emph {et~al.}(2012)\citenamefont
  {Sebastian}, \citenamefont {Ohdaira}, \citenamefont {Kubota}, \citenamefont
  {Pirro}, \citenamefont {BraÌˆcher}, \citenamefont {Vogt}, \citenamefont
  {Serga}, \citenamefont {Naganuma}, \citenamefont {Oogane}, \citenamefont
  {Ando},\ and\ \citenamefont {Hillebrands}}]{Sebastian2012}%
  \BibitemOpen
  \bibfield  {author} {\bibinfo {author} {\bibfnamefont {T.}~\bibnamefont
  {Sebastian}}, \bibinfo {author} {\bibfnamefont {Y.}~\bibnamefont {Ohdaira}},
  \bibinfo {author} {\bibfnamefont {T.}~\bibnamefont {Kubota}}, \bibinfo
  {author} {\bibfnamefont {P.}~\bibnamefont {Pirro}}, \bibinfo {author}
  {\bibfnamefont {T.}~\bibnamefont {BraÌˆcher}}, \bibinfo {author}
  {\bibfnamefont {K.}~\bibnamefont {Vogt}}, \bibinfo {author} {\bibfnamefont
  {A.~A.}\ \bibnamefont {Serga}}, \bibinfo {author} {\bibfnamefont
  {H.}~\bibnamefont {Naganuma}}, \bibinfo {author} {\bibfnamefont
  {M.}~\bibnamefont {Oogane}}, \bibinfo {author} {\bibfnamefont
  {Y.}~\bibnamefont {Ando}}, \ and\ \bibinfo {author} {\bibfnamefont
  {B.}~\bibnamefont {Hillebrands}},\ }\bibfield  {title} {\enquote {\bibinfo
  {title} {Low-damping spin-wave propagation in a micro-structured
  co2mn0.6{F}e0.4{S}i heusler waveguide},}\ }\href {\doibase 10.1063/1.3693391}
  {\bibfield  {journal} {\bibinfo  {journal} {Appl. Phys. Lett.}\ }\textbf
  {\bibinfo {volume} {100}},\ \bibinfo {pages} {112402} (\bibinfo {year}
  {2012})}\BibitemShut {NoStop}%
\bibitem [{\citenamefont {Zhu}\ \emph {et~al.}(2011)\citenamefont {Zhu},
  \citenamefont {Soe}, \citenamefont {McMichael}, \citenamefont {Carey},
  \citenamefont {Maat},\ and\ \citenamefont {Childress}}]{zhu2011}%
  \BibitemOpen
  \bibfield  {author} {\bibinfo {author} {\bibfnamefont {M.}~\bibnamefont
  {Zhu}}, \bibinfo {author} {\bibfnamefont {B.~D.}\ \bibnamefont {Soe}},
  \bibinfo {author} {\bibfnamefont {R.~D.}\ \bibnamefont {McMichael}}, \bibinfo
  {author} {\bibfnamefont {M.~J.}\ \bibnamefont {Carey}}, \bibinfo {author}
  {\bibfnamefont {S.}~\bibnamefont {Maat}}, \ and\ \bibinfo {author}
  {\bibfnamefont {J.~R.}\ \bibnamefont {Childress}},\ }\bibfield  {title}
  {\enquote {\bibinfo {title} {Enhanced magnetization drift velocity and
  current polarization in ({CoFe})[sub 1-x]ge[sub x] alloys},}\ }\href
  {\doibase 10.1063/1.3554755} {\bibfield  {journal} {\bibinfo  {journal}
  {Appl. Phys. Lett.}\ }\textbf {\bibinfo {volume} {98}},\ \bibinfo {pages}
  {072510} (\bibinfo {year} {2011})}\BibitemShut {NoStop}%
\bibitem [{\citenamefont {Yu}\ \emph {et~al.}(2012)\citenamefont {Yu},
  \citenamefont {Huber}, \citenamefont {Schwarze}, \citenamefont {Brandl},
  \citenamefont {Rapp}, \citenamefont {Berberich}, \citenamefont {Duerr},\ and\
  \citenamefont {Grundler}}]{Yu2012}%
  \BibitemOpen
  \bibfield  {author} {\bibinfo {author} {\bibfnamefont {H.}~\bibnamefont
  {Yu}}, \bibinfo {author} {\bibfnamefont {R.}~\bibnamefont {Huber}}, \bibinfo
  {author} {\bibfnamefont {T.}~\bibnamefont {Schwarze}}, \bibinfo {author}
  {\bibfnamefont {F.}~\bibnamefont {Brandl}}, \bibinfo {author} {\bibfnamefont
  {T.}~\bibnamefont {Rapp}}, \bibinfo {author} {\bibfnamefont {P.}~\bibnamefont
  {Berberich}}, \bibinfo {author} {\bibfnamefont {G.}~\bibnamefont {Duerr}}, \
  and\ \bibinfo {author} {\bibfnamefont {D.}~\bibnamefont {Grundler}},\
  }\bibfield  {title} {\enquote {\bibinfo {title} {High propagating velocity of
  spin waves and temperature dependent damping in a {CoFeB} thin film},}\
  }\href {\doibase 10.1063/1.4731273} {\bibfield  {journal} {\bibinfo
  {journal} {Appl. Phys. Lett.}\ }\textbf {\bibinfo {volume} {100}},\ \bibinfo
  {pages} {262412} (\bibinfo {year} {2012})}\BibitemShut {NoStop}%
\bibitem [{\citenamefont {Demidov}\ \emph {et~al.}(2014)\citenamefont
  {Demidov}, \citenamefont {Urazhdin}, \citenamefont {Rinkevich}, \citenamefont
  {Reiss},\ and\ \citenamefont
  {Demokritov}}]{DemidovUrazhdinRinkevichEtAl2014}%
  \BibitemOpen
  \bibfield  {author} {\bibinfo {author} {\bibfnamefont {V.~E.}\ \bibnamefont
  {Demidov}}, \bibinfo {author} {\bibfnamefont {S.}~\bibnamefont {Urazhdin}},
  \bibinfo {author} {\bibfnamefont {A.~B.}\ \bibnamefont {Rinkevich}}, \bibinfo
  {author} {\bibfnamefont {G.}~\bibnamefont {Reiss}}, \ and\ \bibinfo {author}
  {\bibfnamefont {S.~O.}\ \bibnamefont {Demokritov}},\ }\bibfield  {title}
  {\enquote {\bibinfo {title} {Spin hall controlled magnonic
  microwaveguides},}\ }\href {\doibase 10.1063/1.4871519} {\bibfield  {journal}
  {\bibinfo  {journal} {Applied Physics Letters}\ }\textbf {\bibinfo {volume}
  {104}},\ \bibinfo {pages} {152402} (\bibinfo {year} {2014})}\BibitemShut
  {NoStop}%
\bibitem [{\citenamefont {Bhagat}\ and\ \citenamefont
  {Lubitz}(1974)}]{bhagat1974}%
  \BibitemOpen
  \bibfield  {author} {\bibinfo {author} {\bibfnamefont {S.~M.}\ \bibnamefont
  {Bhagat}}\ and\ \bibinfo {author} {\bibfnamefont {P.}~\bibnamefont
  {Lubitz}},\ }\bibfield  {title} {\enquote {\bibinfo {title} {Temperature
  variation of ferromagnetic relaxation in the 3 d transition metals},}\ }\href
  {\doibase 10.1103/physrevb.10.179} {\bibfield  {journal} {\bibinfo  {journal}
  {Phys. Rev. B}\ }\textbf {\bibinfo {volume} {10}},\ \bibinfo {pages}
  {179--185} (\bibinfo {year} {1974})}\BibitemShut {NoStop}%
\bibitem [{\citenamefont {Heinrich}\ and\ \citenamefont
  {Frait}(1966)}]{heinrich1966}%
  \BibitemOpen
  \bibfield  {author} {\bibinfo {author} {\bibfnamefont {B.}~\bibnamefont
  {Heinrich}}\ and\ \bibinfo {author} {\bibfnamefont {Z.}~\bibnamefont
  {Frait}},\ }\bibfield  {title} {\enquote {\bibinfo {title} {Temperature
  dependence of the {FMR} linewidth of iron single-crystal platelets},}\ }\href
  {\doibase 10.1002/pssb.19660160138} {\bibfield  {journal} {\bibinfo
  {journal} {physica status solidi (b)}\ }\textbf {\bibinfo {volume} {16}},\
  \bibinfo {pages} {K11--K14} (\bibinfo {year} {1966})}\BibitemShut {NoStop}%
\bibitem [{\citenamefont {van Bockstal}\ and\ \citenamefont
  {Herlach}(1990)}]{vanbockstal1990}%
  \BibitemOpen
  \bibfield  {author} {\bibinfo {author} {\bibfnamefont {L.}~\bibnamefont {van
  Bockstal}}\ and\ \bibinfo {author} {\bibfnamefont {F.}~\bibnamefont
  {Herlach}},\ }\bibfield  {title} {\enquote {\bibinfo {title} {Ferromagnetic
  relaxation in 3d metals at far infrared frequencies in high magnetic
  fields},}\ }\href {\doibase 10.1088/0953-8984/2/34/012} {\bibfield  {journal}
  {\bibinfo  {journal} {Journal of Physics: Condensed Matter}\ }\textbf
  {\bibinfo {volume} {2}},\ \bibinfo {pages} {7187--7193} (\bibinfo {year}
  {1990})}\BibitemShut {NoStop}%
\bibitem [{\citenamefont {Cochran}\ \emph {et~al.}(1991)\citenamefont
  {Cochran}, \citenamefont {Rudd}, \citenamefont {Muir}, \citenamefont
  {Trayling},\ and\ \citenamefont {Heinrich}}]{cochran1991}%
  \BibitemOpen
  \bibfield  {author} {\bibinfo {author} {\bibfnamefont {J.~F.}\ \bibnamefont
  {Cochran}}, \bibinfo {author} {\bibfnamefont {J.~M.}\ \bibnamefont {Rudd}},
  \bibinfo {author} {\bibfnamefont {W.~B.}\ \bibnamefont {Muir}}, \bibinfo
  {author} {\bibfnamefont {G.}~\bibnamefont {Trayling}}, \ and\ \bibinfo
  {author} {\bibfnamefont {B.}~\bibnamefont {Heinrich}},\ }\bibfield  {title}
  {\enquote {\bibinfo {title} {Temperature dependence of the
  landau{\textendash}lifshitz damping parameter for iron},}\ }\href {\doibase
  10.1063/1.349902} {\bibfield  {journal} {\bibinfo  {journal} {J. Appl.
  Phys.}\ }\textbf {\bibinfo {volume} {70}},\ \bibinfo {pages} {6545} (\bibinfo
  {year} {1991})}\BibitemShut {NoStop}%
\bibitem [{\citenamefont {Urban}, \citenamefont {Woltersdorf},\ and\
  \citenamefont {Heinrich}(2001)}]{urban2001}%
  \BibitemOpen
  \bibfield  {author} {\bibinfo {author} {\bibfnamefont {R.}~\bibnamefont
  {Urban}}, \bibinfo {author} {\bibfnamefont {G.}~\bibnamefont {Woltersdorf}},
  \ and\ \bibinfo {author} {\bibfnamefont {B.}~\bibnamefont {Heinrich}},\
  }\bibfield  {title} {\enquote {\bibinfo {title} {Gilbert damping in single
  and multilayer ultrathin films: Role of interfaces in nonlocal spin
  dynamics},}\ }\href {\doibase 10.1103/physrevlett.87.217204} {\bibfield
  {journal} {\bibinfo  {journal} {Phys. Rev. Lett.}\ }\textbf {\bibinfo
  {volume} {87}} (\bibinfo {year} {2001}),\
  10.1103/physrevlett.87.217204}\BibitemShut {NoStop}%
\bibitem [{\citenamefont {Scheck}, \citenamefont {Cheng},\ and\ \citenamefont
  {Bailey}(2006)}]{ScheckChengBailey2006}%
  \BibitemOpen
  \bibfield  {author} {\bibinfo {author} {\bibfnamefont {C.}~\bibnamefont
  {Scheck}}, \bibinfo {author} {\bibfnamefont {L.}~\bibnamefont {Cheng}}, \
  and\ \bibinfo {author} {\bibfnamefont {W.~E.}\ \bibnamefont {Bailey}},\
  }\bibfield  {title} {\enquote {\bibinfo {title} {Low damping in epitaxial
  sputtered iron films},}\ }\href {\doibase 10.1063/1.2216031} {\bibfield
  {journal} {\bibinfo  {journal} {Applied Physics Letters}\ }\textbf {\bibinfo
  {volume} {88}},\ \bibinfo {pages} {252510} (\bibinfo {year}
  {2006})}\BibitemShut {NoStop}%
\bibitem [{\citenamefont {Yuasa}\ \emph {et~al.}(2004)\citenamefont {Yuasa},
  \citenamefont {Nagahama}, \citenamefont {Fukushima}, \citenamefont {Suzuki},\
  and\ \citenamefont {Ando}}]{yuasa2004}%
  \BibitemOpen
  \bibfield  {author} {\bibinfo {author} {\bibfnamefont {S.}~\bibnamefont
  {Yuasa}}, \bibinfo {author} {\bibfnamefont {T.}~\bibnamefont {Nagahama}},
  \bibinfo {author} {\bibfnamefont {A.}~\bibnamefont {Fukushima}}, \bibinfo
  {author} {\bibfnamefont {Y.}~\bibnamefont {Suzuki}}, \ and\ \bibinfo {author}
  {\bibfnamefont {K.}~\bibnamefont {Ando}},\ }\bibfield  {title} {\enquote
  {\bibinfo {title} {Giant room-temperature magnetoresistance in single-crystal
  fe/{MgO}/fe magnetic tunnel junctions},}\ }\href {\doibase 10.1038/nmat1257}
  {\bibfield  {journal} {\bibinfo  {journal} {Nature Materials}\ }\textbf
  {\bibinfo {volume} {3}},\ \bibinfo {pages} {868--871} (\bibinfo {year}
  {2004})}\BibitemShut {NoStop}%
\bibitem [{\citenamefont {Brockmann}\ \emph {et~al.}(1999)\citenamefont
  {Brockmann}, \citenamefont {Zolfl}, \citenamefont {Miethaner},\ and\
  \citenamefont {Bayreuther}}]{brockmann1999}%
  \BibitemOpen
  \bibfield  {author} {\bibinfo {author} {\bibfnamefont {M.}~\bibnamefont
  {Brockmann}}, \bibinfo {author} {\bibfnamefont {M.}~\bibnamefont {Zolfl}},
  \bibinfo {author} {\bibfnamefont {S.}~\bibnamefont {Miethaner}}, \ and\
  \bibinfo {author} {\bibfnamefont {G.}~\bibnamefont {Bayreuther}},\ }\bibfield
   {title} {\enquote {\bibinfo {title} {In-plane volume and interface magnetic
  anisotropies in epitaxial fe films on {GaAs}(001)},}\ }\href {\doibase
  10.1016/s0304-8853(98)01142-1} {\bibfield  {journal} {\bibinfo  {journal}
  {Journal of Magnetism and Magnetic Materials}\ }\textbf {\bibinfo {volume}
  {198-199}},\ \bibinfo {pages} {384--386} (\bibinfo {year}
  {1999})}\BibitemShut {NoStop}%
\bibitem [{\citenamefont {Vlaminck}\ and\ \citenamefont
  {Bailleul}(2010)}]{VlaminckBailleul2010}%
  \BibitemOpen
  \bibfield  {author} {\bibinfo {author} {\bibfnamefont {V.}~\bibnamefont
  {Vlaminck}}\ and\ \bibinfo {author} {\bibfnamefont {M.}~\bibnamefont
  {Bailleul}},\ }\bibfield  {title} {\enquote {\bibinfo {title} {Spin-wave
  transduction at the submicrometer scale: Experiment and modeling},}\ }\href
  {\doibase 10.1103/physrevb.81.014425} {\bibfield  {journal} {\bibinfo
  {journal} {Physical Review B}\ }\textbf {\bibinfo {volume} {81}},\ \bibinfo
  {pages} {014425} (\bibinfo {year} {2010})}\BibitemShut {NoStop}%
\bibitem [{Note1()}]{Note1}%
  \BibitemOpen
  \bibinfo {note} {The standard expression of MSSW dispersion is modified to
  account for the cubic magnetic anisotropy of Fe and for a small perpendicular
  magnetic anisotropy of magneto-elastic and/or interface origin.}\BibitemShut
  {Stop}%
\bibitem [{\citenamefont {Kalinikos}\ and\ \citenamefont
  {N.}(1986)}]{KalinikosN.1986}%
  \BibitemOpen
  \bibfield  {author} {\bibinfo {author} {\bibfnamefont {B.~A.}\ \bibnamefont
  {Kalinikos}}\ and\ \bibinfo {author} {\bibfnamefont {S.~A.}\ \bibnamefont
  {N.}},\ }\bibfield  {title} {\enquote {\bibinfo {title} {Theory of
  dipole-exchange spin wave spectrum for ferromagnetic films with mixed
  exchange boundary conditions},}\ }\href
  {http://stacks.iop.org/0022-3719/19/i=35/a=014} {\bibfield  {journal}
  {\bibinfo  {journal} {Journal of Physics C: Solid State Physics}\ }\textbf
  {\bibinfo {volume} {19}},\ \bibinfo {pages} {7013} (\bibinfo {year}
  {1986})}\BibitemShut {NoStop}%
\bibitem [{\citenamefont {Craik}(1995)}]{craikbook}%
  \BibitemOpen
  \bibfield  {author} {\bibinfo {author} {\bibfnamefont {D.}~\bibnamefont
  {Craik}},\ }\href@noop {} {\emph {\bibinfo {title} {Magnetism Principles and
  applications}}}\ (\bibinfo  {publisher} {John Wiley and sons},\ \bibinfo
  {year} {1995})\BibitemShut {NoStop}%
\bibitem [{\citenamefont {Gladii}\ \emph
  {et~al.}(2016{\natexlab{a}})\citenamefont {Gladii}, \citenamefont {Collet},
  \citenamefont {Garcia-Hernandez}, \citenamefont {Cheng}, \citenamefont
  {Xavier}, \citenamefont {Bortolotti}, \citenamefont {Cros}, \citenamefont
  {Henry}, \citenamefont {Kim}, \citenamefont {Anane},\ and\ \citenamefont
  {Bailleul}}]{Gladii2016_PyPt}%
  \BibitemOpen
  \bibfield  {author} {\bibinfo {author} {\bibfnamefont {O.}~\bibnamefont
  {Gladii}}, \bibinfo {author} {\bibfnamefont {M.}~\bibnamefont {Collet}},
  \bibinfo {author} {\bibfnamefont {K.}~\bibnamefont {Garcia-Hernandez}},
  \bibinfo {author} {\bibfnamefont {C.}~\bibnamefont {Cheng}}, \bibinfo
  {author} {\bibfnamefont {S.}~\bibnamefont {Xavier}}, \bibinfo {author}
  {\bibfnamefont {P.}~\bibnamefont {Bortolotti}}, \bibinfo {author}
  {\bibfnamefont {V.}~\bibnamefont {Cros}}, \bibinfo {author} {\bibfnamefont
  {Y.}~\bibnamefont {Henry}}, \bibinfo {author} {\bibfnamefont {J.-V.}\
  \bibnamefont {Kim}}, \bibinfo {author} {\bibfnamefont {A.}~\bibnamefont
  {Anane}}, \ and\ \bibinfo {author} {\bibfnamefont {M.}~\bibnamefont
  {Bailleul}},\ }\bibfield  {title} {\enquote {\bibinfo {title} {Spin wave
  amplification using the spin hall effect in permalloy/platinum bilayers},}\
  }\href {\doibase 10.1063/1.4952447} {\bibfield  {journal} {\bibinfo
  {journal} {Appl. Phys. Lett.}\ }\textbf {\bibinfo {volume} {108}},\ \bibinfo
  {pages} {202407} (\bibinfo {year} {2016}{\natexlab{a}})}\BibitemShut
  {NoStop}%
\bibitem [{\citenamefont {Chang}\ \emph {et~al.}(2014)\citenamefont {Chang},
  \citenamefont {Kostylev}, \citenamefont {Ivanov}, \citenamefont {Ding},\ and\
  \citenamefont {Adeyeye}}]{ChangKostylevIvanovEtAl2014}%
  \BibitemOpen
  \bibfield  {author} {\bibinfo {author} {\bibfnamefont {C.~S.}\ \bibnamefont
  {Chang}}, \bibinfo {author} {\bibfnamefont {M.}~\bibnamefont {Kostylev}},
  \bibinfo {author} {\bibfnamefont {E.}~\bibnamefont {Ivanov}}, \bibinfo
  {author} {\bibfnamefont {J.}~\bibnamefont {Ding}}, \ and\ \bibinfo {author}
  {\bibfnamefont {A.~O.}\ \bibnamefont {Adeyeye}},\ }\bibfield  {title}
  {\enquote {\bibinfo {title} {The phase accumulation and antenna near field of
  microscopic propagating spin wave devices},}\ }\href {\doibase
  10.1063/1.4863078} {\bibfield  {journal} {\bibinfo  {journal} {Appl. Phys.
  Lett.}\ }\textbf {\bibinfo {volume} {104}},\ \bibinfo {pages} {032408}
  (\bibinfo {year} {2014})}\BibitemShut {NoStop}%
\bibitem [{\citenamefont {Schneider}\ \emph {et~al.}(2008)\citenamefont
  {Schneider}, \citenamefont {Serga}, \citenamefont {Neumann}, \citenamefont
  {Hillebrands},\ and\ \citenamefont
  {Kostylev}}]{SchneiderSergaNeumannEtAl2008}%
  \BibitemOpen
  \bibfield  {author} {\bibinfo {author} {\bibfnamefont {T.}~\bibnamefont
  {Schneider}}, \bibinfo {author} {\bibfnamefont {A.~A.}\ \bibnamefont
  {Serga}}, \bibinfo {author} {\bibfnamefont {T.}~\bibnamefont {Neumann}},
  \bibinfo {author} {\bibfnamefont {B.}~\bibnamefont {Hillebrands}}, \ and\
  \bibinfo {author} {\bibfnamefont {M.~P.}\ \bibnamefont {Kostylev}},\
  }\bibfield  {title} {\enquote {\bibinfo {title} {Phase reciprocity of
  spin-wave excitation by a microstrip antenna},}\ }\href {\doibase
  10.1103/physrevb.77.214411} {\bibfield  {journal} {\bibinfo  {journal}
  {Physical Review B}\ }\textbf {\bibinfo {volume} {77}},\ \bibinfo {pages}
  {214411} (\bibinfo {year} {2008})}\BibitemShut {NoStop}%
\bibitem [{\citenamefont {Gladii}\ \emph
  {et~al.}(2016{\natexlab{b}})\citenamefont {Gladii}, \citenamefont {Haidar},
  \citenamefont {Henry}, \citenamefont {Kostylev},\ and\ \citenamefont
  {Bailleul}}]{GladiiHaidarHenryEtAl2016}%
  \BibitemOpen
  \bibfield  {author} {\bibinfo {author} {\bibfnamefont {O.}~\bibnamefont
  {Gladii}}, \bibinfo {author} {\bibfnamefont {M.}~\bibnamefont {Haidar}},
  \bibinfo {author} {\bibfnamefont {Y.}~\bibnamefont {Henry}}, \bibinfo
  {author} {\bibfnamefont {M.}~\bibnamefont {Kostylev}}, \ and\ \bibinfo
  {author} {\bibfnamefont {M.}~\bibnamefont {Bailleul}},\ }\bibfield  {title}
  {\enquote {\bibinfo {title} {Frequency nonreciprocity of surface spin wave in
  permalloy thin films},}\ }\href {\doibase 10.1103/physrevb.93.054430}
  {\bibfield  {journal} {\bibinfo  {journal} {Physical Review B}\ }\textbf
  {\bibinfo {volume} {93}},\ \bibinfo {pages} {054430} (\bibinfo {year}
  {2016}{\natexlab{b}})}\BibitemShut {NoStop}%
\bibitem [{\citenamefont {Haidar}\ and\ \citenamefont
  {Bailleul}(2013)}]{Haidar2013}%
  \BibitemOpen
  \bibfield  {author} {\bibinfo {author} {\bibfnamefont {M.}~\bibnamefont
  {Haidar}}\ and\ \bibinfo {author} {\bibfnamefont {M.}~\bibnamefont
  {Bailleul}},\ }\bibfield  {title} {\enquote {\bibinfo {title} {Thickness
  dependence of degree of spin polarization of electrical current in permalloy
  thin films},}\ }\href {\doibase 10.1103/physrevb.88.054417} {\bibfield
  {journal} {\bibinfo  {journal} {Phys. Rev. B}\ }\textbf {\bibinfo {volume}
  {88}} (\bibinfo {year} {2013}),\ 10.1103/physrevb.88.054417}\BibitemShut
  {NoStop}%
\bibitem [{\citenamefont {Haidar}\ \emph {et~al.}(2014)\citenamefont {Haidar},
  \citenamefont {Bailleul}, \citenamefont {Kostylev},\ and\ \citenamefont
  {Lao}}]{HaidarBailleulKostylevEtAl2014}%
  \BibitemOpen
  \bibfield  {author} {\bibinfo {author} {\bibfnamefont {M.}~\bibnamefont
  {Haidar}}, \bibinfo {author} {\bibfnamefont {M.}~\bibnamefont {Bailleul}},
  \bibinfo {author} {\bibfnamefont {M.~P.}\ \bibnamefont {Kostylev}}, \ and\
  \bibinfo {author} {\bibfnamefont {Y.}~\bibnamefont {Lao}},\ }\bibfield
  {title} {\enquote {\bibinfo {title} {Nonreciprocal {Oersted} field
  contribution to the current-induced frequency shift of magnetostatic surface
  waves},}\ }\href {\doibase 10.1103/physrevb.89.094426} {\bibfield  {journal}
  {\bibinfo  {journal} {Physical Review B}\ }\textbf {\bibinfo {volume} {89}},\
  \bibinfo {pages} {094426} (\bibinfo {year} {2014})}\BibitemShut {NoStop}%
\bibitem [{\citenamefont {Zhu}, \citenamefont {Dennis},\ and\ \citenamefont
  {McMichael}(2010)}]{Zhu2010}%
  \BibitemOpen
  \bibfield  {author} {\bibinfo {author} {\bibfnamefont {M.}~\bibnamefont
  {Zhu}}, \bibinfo {author} {\bibfnamefont {C.~L.}\ \bibnamefont {Dennis}}, \
  and\ \bibinfo {author} {\bibfnamefont {R.~D.}\ \bibnamefont {McMichael}},\
  }\bibfield  {title} {\enquote {\bibinfo {title} {Temperature dependence of
  magnetization drift velocity and current polarization in ni 80~{F}e 20~{b}y
  spin-wave {Doppler} measurements},}\ }\href {\doibase
  10.1103/physrevb.81.140407} {\bibfield  {journal} {\bibinfo  {journal} {Phys.
  Rev. B}\ }\textbf {\bibinfo {volume} {81}} (\bibinfo {year} {2010}),\
  10.1103/physrevb.81.140407}\BibitemShut {NoStop}%
\bibitem [{\citenamefont {Sekiguchi}\ \emph {et~al.}(2012)\citenamefont
  {Sekiguchi}, \citenamefont {Yamada}, \citenamefont {Seo}, \citenamefont
  {Lee}, \citenamefont {Chiba}, \citenamefont {Kobayashi},\ and\ \citenamefont
  {Ono}}]{Sekiguchi2012}%
  \BibitemOpen
  \bibfield  {author} {\bibinfo {author} {\bibfnamefont {K.}~\bibnamefont
  {Sekiguchi}}, \bibinfo {author} {\bibfnamefont {K.}~\bibnamefont {Yamada}},
  \bibinfo {author} {\bibfnamefont {S.-M.}\ \bibnamefont {Seo}}, \bibinfo
  {author} {\bibfnamefont {K.-J.}\ \bibnamefont {Lee}}, \bibinfo {author}
  {\bibfnamefont {D.}~\bibnamefont {Chiba}}, \bibinfo {author} {\bibfnamefont
  {K.}~\bibnamefont {Kobayashi}}, \ and\ \bibinfo {author} {\bibfnamefont
  {T.}~\bibnamefont {Ono}},\ }\bibfield  {title} {\enquote {\bibinfo {title}
  {Time-domain measurement of current-induced spin wave dynamics},}\ }\href
  {\doibase 10.1103/physrevlett.108.017203} {\bibfield  {journal} {\bibinfo
  {journal} {Phys. Rev. Lett.}\ }\textbf {\bibinfo {volume} {108}} (\bibinfo
  {year} {2012}),\ 10.1103/physrevlett.108.017203}\BibitemShut {NoStop}%
\bibitem [{\citenamefont {Chauleau}\ \emph {et~al.}(2014)\citenamefont
  {Chauleau}, \citenamefont {Bauer}, \citenamefont {Körner}, \citenamefont
  {Stigloher}, \citenamefont {Härtinger}, \citenamefont {Woltersdorf},\ and\
  \citenamefont {Back}}]{ChauleauBauerKoernerEtAl2014}%
  \BibitemOpen
  \bibfield  {author} {\bibinfo {author} {\bibfnamefont {J.-Y.}\ \bibnamefont
  {Chauleau}}, \bibinfo {author} {\bibfnamefont {H.~G.}\ \bibnamefont {Bauer}},
  \bibinfo {author} {\bibfnamefont {H.~S.}\ \bibnamefont {Körner}}, \bibinfo
  {author} {\bibfnamefont {J.}~\bibnamefont {Stigloher}}, \bibinfo {author}
  {\bibfnamefont {M.}~\bibnamefont {Härtinger}}, \bibinfo {author}
  {\bibfnamefont {G.}~\bibnamefont {Woltersdorf}}, \ and\ \bibinfo {author}
  {\bibfnamefont {C.~H.}\ \bibnamefont {Back}},\ }\bibfield  {title} {\enquote
  {\bibinfo {title} {Self-consistent determination of the key spin-transfer
  torque parameters from spin-wave {Doppler} experiments},}\ }\href {\doibase
  10.1103/physrevb.89.020403} {\bibfield  {journal} {\bibinfo  {journal} {Phys.
  Rev. B}\ }\textbf {\bibinfo {volume} {89}} (\bibinfo {year} {2014}),\
  10.1103/physrevb.89.020403}\BibitemShut {NoStop}%
\bibitem [{\citenamefont {Campbell}\ and\ \citenamefont
  {Fert}(1982)}]{CampbellFertchapter}%
  \BibitemOpen
  \bibfield  {author} {\bibinfo {author} {\bibfnamefont {I.}~\bibnamefont
  {Campbell}}\ and\ \bibinfo {author} {\bibfnamefont {A.}~\bibnamefont
  {Fert}},\ }\enquote {\bibinfo {title} {Ferromagnetic materials, vol. 3},}\ \
  (\bibinfo  {publisher} {North-Holland},\ \bibinfo {year} {1982})\ Chap.\
  \bibinfo {chapter} {Transport Properties of Ferromagnets}, pp.\ \bibinfo
  {pages} {747--804},\ \bibinfo {note} {e.P. Wohlfarth, Ed.}\BibitemShut
  {Stop}%
\bibitem [{\citenamefont {Mertig}(1999)}]{Mertig1999}%
  \BibitemOpen
  \bibfield  {author} {\bibinfo {author} {\bibfnamefont {I.}~\bibnamefont
  {Mertig}},\ }\bibfield  {title} {\enquote {\bibinfo {title} {Transport
  properties of dilute alloys},}\ }\href {\doibase 10.1088/0034-4885/62/2/004}
  {\bibfield  {journal} {\bibinfo  {journal} {Rep. Prog. Phys.}\ }\textbf
  {\bibinfo {volume} {62}},\ \bibinfo {pages} {237--276} (\bibinfo {year}
  {1999})}\BibitemShut {NoStop}%
\bibitem [{\citenamefont {Gurney}\ \emph {et~al.}(1993)\citenamefont {Gurney},
  \citenamefont {Speriosu}, \citenamefont {Nozieres}, \citenamefont {Lefakis},
  \citenamefont {Wilhoit},\ and\ \citenamefont
  {Need}}]{GurneySperiosuNozieresEtAl1993}%
  \BibitemOpen
  \bibfield  {author} {\bibinfo {author} {\bibfnamefont {B.~A.}\ \bibnamefont
  {Gurney}}, \bibinfo {author} {\bibfnamefont {V.~S.}\ \bibnamefont
  {Speriosu}}, \bibinfo {author} {\bibfnamefont {J.-P.}\ \bibnamefont
  {Nozieres}}, \bibinfo {author} {\bibfnamefont {H.}~\bibnamefont {Lefakis}},
  \bibinfo {author} {\bibfnamefont {D.~R.}\ \bibnamefont {Wilhoit}}, \ and\
  \bibinfo {author} {\bibfnamefont {O.~U.}\ \bibnamefont {Need}},\ }\bibfield
  {title} {\enquote {\bibinfo {title} {Direct measurement of spin-dependent
  conduction-electron mean free paths in ferromagnetic metals},}\ }\href
  {\doibase 10.1103/physrevlett.71.4023} {\bibfield  {journal} {\bibinfo
  {journal} {Physical Review Letters}\ }\textbf {\bibinfo {volume} {71}},\
  \bibinfo {pages} {4023--4026} (\bibinfo {year} {1993})}\BibitemShut {NoStop}%
\bibitem [{\citenamefont {Fert}\ and\ \citenamefont
  {Campbell}(1976)}]{FertCampbell1976}%
  \BibitemOpen
  \bibfield  {author} {\bibinfo {author} {\bibfnamefont {A.}~\bibnamefont
  {Fert}}\ and\ \bibinfo {author} {\bibfnamefont {I.~A.}\ \bibnamefont
  {Campbell}},\ }\bibfield  {title} {\enquote {\bibinfo {title} {Electrical
  resistivity of ferromagnetic nickel and iron based alloys},}\ }\href
  {\doibase 10.1088/0305-4608/6/5/025} {\bibfield  {journal} {\bibinfo
  {journal} {Journal of Physics F: Metal Physics}\ }\textbf {\bibinfo {volume}
  {6}},\ \bibinfo {pages} {849--871} (\bibinfo {year} {1976})}\BibitemShut
  {NoStop}%
\bibitem [{\citenamefont {Loegel}\ and\ \citenamefont
  {Gautier}(1971)}]{LoegelGautier1971}%
  \BibitemOpen
  \bibfield  {author} {\bibinfo {author} {\bibfnamefont {B.}~\bibnamefont
  {Loegel}}\ and\ \bibinfo {author} {\bibfnamefont {F.}~\bibnamefont
  {Gautier}},\ }\bibfield  {title} {\enquote {\bibinfo {title} {Origine de la
  resistivite dans le cobalt et ses alliages dilues},}\ }\href {\doibase
  10.1016/s0022-3697(71)80364-5} {\bibfield  {journal} {\bibinfo  {journal}
  {Journal of Physics and Chemistry of Solids}\ }\textbf {\bibinfo {volume}
  {32}},\ \bibinfo {pages} {2723--2735} (\bibinfo {year} {1971})}\BibitemShut
  {NoStop}%
\bibitem [{\citenamefont {Raquet}\ \emph {et~al.}(2002)\citenamefont {Raquet},
  \citenamefont {Viret}, \citenamefont {Sondergard}, \citenamefont {Cespedes},\
  and\ \citenamefont {Mamy}}]{RaquetViretSondergardEtAl2002}%
  \BibitemOpen
  \bibfield  {author} {\bibinfo {author} {\bibfnamefont {B.}~\bibnamefont
  {Raquet}}, \bibinfo {author} {\bibfnamefont {M.}~\bibnamefont {Viret}},
  \bibinfo {author} {\bibfnamefont {E.}~\bibnamefont {Sondergard}}, \bibinfo
  {author} {\bibfnamefont {O.}~\bibnamefont {Cespedes}}, \ and\ \bibinfo
  {author} {\bibfnamefont {R.}~\bibnamefont {Mamy}},\ }\bibfield  {title}
  {\enquote {\bibinfo {title} {Electron-magnon scattering and magnetic
  resistivity in 3 d ferromagnets},}\ }\href {\doibase
  10.1103/physrevb.66.024433} {\bibfield  {journal} {\bibinfo  {journal}
  {Physical Review B}\ }\textbf {\bibinfo {volume} {66}},\ \bibinfo {pages}
  {024433} (\bibinfo {year} {2002})}\BibitemShut {NoStop}%
\bibitem [{\citenamefont {Liu}\ \emph {et~al.}(2015)\citenamefont {Liu},
  \citenamefont {Yuan}, \citenamefont {Wesselink}, \citenamefont {Starikov},
  \citenamefont {van Schilfgaarde},\ and\ \citenamefont
  {Kelly}}]{LiuYuanWesselinkEtAl2015}%
  \BibitemOpen
  \bibfield  {author} {\bibinfo {author} {\bibfnamefont {Y.}~\bibnamefont
  {Liu}}, \bibinfo {author} {\bibfnamefont {Z.}~\bibnamefont {Yuan}}, \bibinfo
  {author} {\bibfnamefont {R.~J.~H.}\ \bibnamefont {Wesselink}}, \bibinfo
  {author} {\bibfnamefont {A.~A.}\ \bibnamefont {Starikov}}, \bibinfo {author}
  {\bibfnamefont {M.}~\bibnamefont {van Schilfgaarde}}, \ and\ \bibinfo
  {author} {\bibfnamefont {P.~J.}\ \bibnamefont {Kelly}},\ }\bibfield  {title}
  {\enquote {\bibinfo {title} {Direct method for calculating
  temperature-dependent transport properties},}\ }\href {\doibase
  10.1103/physrevb.91.220405} {\bibfield  {journal} {\bibinfo  {journal} {Phys.
  Rev. B}\ }\textbf {\bibinfo {volume} {91}} (\bibinfo {year} {2015}),\
  10.1103/physrevb.91.220405}\BibitemShut {NoStop}%
\bibitem [{\citenamefont {Ebert}\ \emph {et~al.}(2015)\citenamefont {Ebert},
  \citenamefont {Mankovsky}, \citenamefont {Chadova}, \citenamefont {Polesya},
  \citenamefont {Min{\'{a}}r},\ and\ \citenamefont
  {Ködderitzsch}}]{EbertMankovskyChadovaEtAl2015}%
  \BibitemOpen
  \bibfield  {author} {\bibinfo {author} {\bibfnamefont {H.}~\bibnamefont
  {Ebert}}, \bibinfo {author} {\bibfnamefont {S.}~\bibnamefont {Mankovsky}},
  \bibinfo {author} {\bibfnamefont {K.}~\bibnamefont {Chadova}}, \bibinfo
  {author} {\bibfnamefont {S.}~\bibnamefont {Polesya}}, \bibinfo {author}
  {\bibfnamefont {J.}~\bibnamefont {Min{\'{a}}r}}, \ and\ \bibinfo {author}
  {\bibfnamefont {D.}~\bibnamefont {Ködderitzsch}},\ }\bibfield  {title}
  {\enquote {\bibinfo {title} {Calculating linear-response functions for finite
  temperatures on the basis of the alloy analogy model},}\ }\href {\doibase
  10.1103/physrevb.91.165132} {\bibfield  {journal} {\bibinfo  {journal} {Phys.
  Rev. B}\ }\textbf {\bibinfo {volume} {91}} (\bibinfo {year} {2015}),\
  10.1103/physrevb.91.165132}\BibitemShut {NoStop}%
\bibitem [{\citenamefont {Savrasov}\ and\ \citenamefont
  {Savrasov}(1996)}]{SavrasovSavrasov1996}%
  \BibitemOpen
  \bibfield  {author} {\bibinfo {author} {\bibfnamefont {S.~Y.}\ \bibnamefont
  {Savrasov}}\ and\ \bibinfo {author} {\bibfnamefont {D.~Y.}\ \bibnamefont
  {Savrasov}},\ }\bibfield  {title} {\enquote {\bibinfo {title}
  {Electron-phonon interactions and related physical properties of metals from
  linear-response theory},}\ }\href {\doibase 10.1103/physrevb.54.16487}
  {\bibfield  {journal} {\bibinfo  {journal} {Physical Review B}\ }\textbf
  {\bibinfo {volume} {54}},\ \bibinfo {pages} {16487--16501} (\bibinfo {year}
  {1996})}\BibitemShut {NoStop}%
\bibitem [{\citenamefont {Verstraete}(2013)}]{Verstraete2013}%
  \BibitemOpen
  \bibfield  {author} {\bibinfo {author} {\bibfnamefont {M.~J.}\ \bibnamefont
  {Verstraete}},\ }\bibfield  {title} {\enquote {\bibinfo {title} {Ab initio
  calculation of spin-dependent electronâ€“phonon coupling in iron and
  cobalt},}\ }\href {\doibase 10.1088/0953-8984/25/13/136001} {\bibfield
  {journal} {\bibinfo  {journal} {J. Phys.: Condens. Matter}\ }\textbf
  {\bibinfo {volume} {25}},\ \bibinfo {pages} {136001} (\bibinfo {year}
  {2013})}\BibitemShut {NoStop}%
\bibitem [{\citenamefont {Zhu}\ \emph {et~al.}(2008)\citenamefont {Zhu},
  \citenamefont {Zhang}, \citenamefont {Xu}, \citenamefont {Chen},
  \citenamefont {Wu}, \citenamefont {Zhang},\ and\ \citenamefont
  {Zhang}}]{ZhuZhangXuEtAl2008}%
  \BibitemOpen
  \bibfield  {author} {\bibinfo {author} {\bibfnamefont {Z.~Y.}\ \bibnamefont
  {Zhu}}, \bibinfo {author} {\bibfnamefont {H.~W.}\ \bibnamefont {Zhang}},
  \bibinfo {author} {\bibfnamefont {S.~F.}\ \bibnamefont {Xu}}, \bibinfo
  {author} {\bibfnamefont {J.~L.}\ \bibnamefont {Chen}}, \bibinfo {author}
  {\bibfnamefont {G.~H.}\ \bibnamefont {Wu}}, \bibinfo {author} {\bibfnamefont
  {B.}~\bibnamefont {Zhang}}, \ and\ \bibinfo {author} {\bibfnamefont {X.~X.}\
  \bibnamefont {Zhang}},\ }\bibfield  {title} {\enquote {\bibinfo {title}
  {Intrinsic anisotropy of degree of transport spin polarization in typical
  ferromagnets},}\ }\href {\doibase 10.1088/0953-8984/20/27/275245} {\bibfield
  {journal} {\bibinfo  {journal} {J. Phys.: Condens. Matter}\ }\textbf
  {\bibinfo {volume} {20}},\ \bibinfo {pages} {275245} (\bibinfo {year}
  {2008})}\BibitemShut {NoStop}%
\bibitem [{\citenamefont {Soulen~Jr.}(1998)}]{SoulenJr.1998}%
  \BibitemOpen
  \bibfield  {author} {\bibinfo {author} {\bibfnamefont {R.~J.}\ \bibnamefont
  {Soulen~Jr.}},\ }\bibfield  {title} {\enquote {\bibinfo {title} {Measuring
  the spin polarization of a metal with a superconducting point contact},}\
  }\href {\doibase 10.1126/science.282.5386.85} {\bibfield  {journal} {\bibinfo
   {journal} {Science}\ }\textbf {\bibinfo {volume} {282}},\ \bibinfo {pages}
  {85--88} (\bibinfo {year} {1998})}\BibitemShut {NoStop}%
\bibitem [{\citenamefont {Scheck}\ \emph {et~al.}(2007)\citenamefont {Scheck},
  \citenamefont {Cheng}, \citenamefont {Barsukov}, \citenamefont {Frait},\ and\
  \citenamefont {Bailey}}]{ScheckChengBarsukovEtAl2007}%
  \BibitemOpen
  \bibfield  {author} {\bibinfo {author} {\bibfnamefont {C.}~\bibnamefont
  {Scheck}}, \bibinfo {author} {\bibfnamefont {L.}~\bibnamefont {Cheng}},
  \bibinfo {author} {\bibfnamefont {I.}~\bibnamefont {Barsukov}}, \bibinfo
  {author} {\bibfnamefont {Z.}~\bibnamefont {Frait}}, \ and\ \bibinfo {author}
  {\bibfnamefont {W.~E.}\ \bibnamefont {Bailey}},\ }\bibfield  {title}
  {\enquote {\bibinfo {title} {Low relaxation rate in epitaxial vanadium-doped
  ultrathin iron films},}\ }\href {\doibase 10.1103/physrevlett.98.117601}
  {\bibfield  {journal} {\bibinfo  {journal} {Phys. Rev. Lett.}\ }\textbf
  {\bibinfo {volume} {98}} (\bibinfo {year} {2007}),\
  10.1103/physrevlett.98.117601}\BibitemShut {NoStop}%
\bibitem [{\citenamefont {Devolder}\ \emph {et~al.}(2013)\citenamefont
  {Devolder}, \citenamefont {Tahmasebi}, \citenamefont {Eimer}, \citenamefont
  {Hauet},\ and\ \citenamefont {Andrieu}}]{devolder2013}%
  \BibitemOpen
  \bibfield  {author} {\bibinfo {author} {\bibfnamefont {T.}~\bibnamefont
  {Devolder}}, \bibinfo {author} {\bibfnamefont {T.}~\bibnamefont {Tahmasebi}},
  \bibinfo {author} {\bibfnamefont {S.}~\bibnamefont {Eimer}}, \bibinfo
  {author} {\bibfnamefont {T.}~\bibnamefont {Hauet}}, \ and\ \bibinfo {author}
  {\bibfnamefont {S.}~\bibnamefont {Andrieu}},\ }\bibfield  {title} {\enquote
  {\bibinfo {title} {Compositional dependence of the magnetic properties of
  epitaxial {FeV}/{MgO} thin films},}\ }\href {\doibase 10.1063/1.4845375}
  {\bibfield  {journal} {\bibinfo  {journal} {Appl. Phys. Lett.}\ }\textbf
  {\bibinfo {volume} {103}},\ \bibinfo {pages} {242410} (\bibinfo {year}
  {2013})}\BibitemShut {NoStop}%
\bibitem [{\citenamefont {Schoen}\ \emph {et~al.}(2016)\citenamefont {Schoen},
  \citenamefont {Thonig}, \citenamefont {Schneider}, \citenamefont {Silva},
  \citenamefont {Nembach}, \citenamefont {Eriksson}, \citenamefont {Karis},\
  and\ \citenamefont {Shaw}}]{SchoenThonigSchneiderEtAl2016}%
  \BibitemOpen
  \bibfield  {author} {\bibinfo {author} {\bibfnamefont {M.~A.~W.}\
  \bibnamefont {Schoen}}, \bibinfo {author} {\bibfnamefont {D.}~\bibnamefont
  {Thonig}}, \bibinfo {author} {\bibfnamefont {M.~L.}\ \bibnamefont
  {Schneider}}, \bibinfo {author} {\bibfnamefont {T.~J.}\ \bibnamefont
  {Silva}}, \bibinfo {author} {\bibfnamefont {H.~T.}\ \bibnamefont {Nembach}},
  \bibinfo {author} {\bibfnamefont {O.}~\bibnamefont {Eriksson}}, \bibinfo
  {author} {\bibfnamefont {O.}~\bibnamefont {Karis}}, \ and\ \bibinfo {author}
  {\bibfnamefont {J.~M.}\ \bibnamefont {Shaw}},\ }\bibfield  {title} {\enquote
  {\bibinfo {title} {Ultra-low magnetic damping of a metallic~ferromagnet},}\
  }\href {\doibase 10.1038/nphys3770} {\bibfield  {journal} {\bibinfo
  {journal} {Nat Phys}\ } (\bibinfo {year} {2016}),\
  10.1038/nphys3770}\BibitemShut {NoStop}%
\bibitem [{\citenamefont {Zahnd}\ \emph {et~al.}(2015)\citenamefont {Zahnd},
  \citenamefont {Vila}, \citenamefont {Pham}, \citenamefont {Marty},
  \citenamefont {Laczkowski}, \citenamefont {Torres}, \citenamefont
  {Beign{\'{e}}}, \citenamefont {Vergnaud}, \citenamefont {Jamet},\ and\
  \citenamefont {Attan{\'{e}}}}]{ZahndVilaPhamEtAl2015}%
  \BibitemOpen
  \bibfield  {author} {\bibinfo {author} {\bibfnamefont {G.}~\bibnamefont
  {Zahnd}}, \bibinfo {author} {\bibfnamefont {L.}~\bibnamefont {Vila}},
  \bibinfo {author} {\bibfnamefont {T.~V.}\ \bibnamefont {Pham}}, \bibinfo
  {author} {\bibfnamefont {A.}~\bibnamefont {Marty}}, \bibinfo {author}
  {\bibfnamefont {P.}~\bibnamefont {Laczkowski}}, \bibinfo {author}
  {\bibfnamefont {W.~S.}\ \bibnamefont {Torres}}, \bibinfo {author}
  {\bibfnamefont {C.}~\bibnamefont {Beign{\'{e}}}}, \bibinfo {author}
  {\bibfnamefont {C.}~\bibnamefont {Vergnaud}}, \bibinfo {author}
  {\bibfnamefont {M.}~\bibnamefont {Jamet}}, \ and\ \bibinfo {author}
  {\bibfnamefont {J.-P.}\ \bibnamefont {Attan{\'{e}}}},\ }\bibfield  {title}
  {\enquote {\bibinfo {title} {Comparison of the use of {NiFe} and {CoFe} as
  electrodes for metallic lateral spin valves},}\ }\href {\doibase
  10.1088/0957-4484/27/3/035201} {\bibfield  {journal} {\bibinfo  {journal}
  {Nanotechnology}\ }\textbf {\bibinfo {volume} {27}},\ \bibinfo {pages}
  {035201} (\bibinfo {year} {2015})}\BibitemShut {NoStop}%
\end{thebibliography}

%

\end{document}